\documentclass[a4paper,12pt]{article}
\pdfoutput=1 
\usepackage{jcappub} 
\usepackage[T1]{fontenc} 
\title{Cosmological Constant Problem and
Renormalized Vacuum Energy Density in Curved Background}
\author[a,b]{Kazunori Kohri,}
\author[b]{Hiroki Matsui}
\affiliation[a]{Theory Center, IPNS, KEK,  Tsukuba 305-0801, Ibaraki, Japan}
\affiliation[b]{The Graduate University of Advanced Studies (Sokendai), Tsukuba 305-0801, Ibaraki, Japan}
\emailAdd{kohri@post.kek.jp} \emailAdd{matshiro@post.kek.jp}

\abstract{
The current vacuum energy density observed as dark energy ${ \rho }_{ \rm dark }\simeq 2.5\times10^{-47}\ {\rm GeV^{4}}$
is unacceptably small compared with any other scales. 
Therefore, we encounter serious fine-tuning problem and theoretical difficulty to derive the dark energy.
However, the theoretically attractive scenario has been proposed and discussed in literature:
In terms of the renormalization-group (RG) running of the cosmological constant,
the vacuum energy density can be expressed as ${ \rho }_{ \rm vacuum }\simeq m^{2}H^{2}$ where $m$ is 
the mass of the scalar field and rather dynamical in curved spacetime.
However, there has been no rigorous proof to derive this expression
and there are some criticisms about the physical interpretation of the RG running cosmological constant.
In the present paper, we revisit the RG running effects of the cosmological constant and
investigate the renormalized vacuum energy density in curved spacetime.  
We demonstrate that the vacuum energy density described by ${ \rho }_{ \rm vacuum }\simeq m^{2}H^{2}$ appears as quantum effects
of the curved background rather than the running effects of cosmological constant.
Comparing to cosmological observational data, we obtain an upper bound on the mass of the scalar fields to be
smaller than the Planck mass, $m  \lesssim M_{\rm Pl}$.}

\begin{document}
\maketitle
\flushbottom
\newcommand{\Slash}[1]{{\ooalign{\hfil/\hfil\crcr$#1$}}}
\section{Introduction}
\label{sec:intro}
From the modern perspective of standard quantum field theory (QFT),
the cosmological constant (CC) term in Einstein's equations can be
interpreted as a quantum vacuum energy density.  As well-known facts,
however, there is an unacceptably huge discrepancy between the
theoretical prediction of standard QFT and the observed dark energy
obtained from current cosmological data
\cite{Knop:2003iy,Riess:2004nr,Spergel:2006hy,0067-0049-192-2-18,Ade:2013zuv}.
This is known as the cosmological constant problem or the dark energy
problem, which is recognized as one of the most profound and notorious
puzzles in theoretical physics
\cite{Weinberg:1988cp,Sahni:1999gb,Carroll:2000fy,Padmanabhan:2002ji,
  Nojiri:2006ri,Li:2011sd,Wang:2016och,Martin:2012bt,Peebles:2002gy,Copeland:2006wr}.
Despite enormous theoretical efforts, unfortunately there has been no
satisfactory scenario to solve the problem about the desperate
fine-tuning of the vacuum energy density in any realistic ways and
derive the dark energy scale,
${ \rho }_{ \rm dark }\simeq 2.5\times10^{-47}\ {\rm GeV^{4}}$, which
is far from the energy scales of the electroweak theory, the grand
unified theory (GUT) and the Planck scale physics.

In the framework of the standard model (SM), we can write down the
vacuum energy density as follows:
\begin{equation}
{ \rho  }_{\rm vacuum  }={ \rho  }_{ \Lambda  }+{ \rho  }_{ \rm  EW }+{ \rho  }_{ \rm  QCD }
+\sum _{ i }^{  }{  \frac { n_{i}{ m }_{i}^{ 4 }}{ 64{ \pi  }^{ 2 } }
\left( \ln {  \frac { { m }_{i}^{ 4 }  }{ { \mu  }^{ 2 } } } -\frac{3}{2} \right)}+{ \rho  }_{ \rm  UV}+\cdots,
\end{equation}
where ${ \rho }_{ \Lambda }$ means just a cosmological constant in
Einstein's equations.  ${ \rho }_{ \rm EW }\sim 10^{8}\ {\rm GeV^{4}}$
and ${ \rho }_{ \rm QCD }\sim 10^{-2}\ {\rm GeV^{4}}$ are the
classical vacuum energies of the electroweak symmetry breaking and the
chiral symmetry breaking, respectively. The logarithm terms describe
the quantum corrections of the vacuum energy density where $n_{i}$ and
$m_{i}$ express the number of degrees of freedom and the mass of the
SM particle $i$, respectively. $\mu$ is the renormalization scale.
${ \rho }_{ \rm UV }$ expresses a possible vacuum energy density
expected in ultraviolet (UV) theories.  However, the current physical
value of the vacuum energy density observed as the dark energy,
${ \rho }_{ \rm dark }\simeq 2.5\times10^{-47}\ {\rm GeV^{4}}$ is
unacceptably and extremely smaller than the theoretical predictions of
those vacuum energy densities.  Therefore, we encounter the serious
fine-tuning of the vacuum energy and the theoretical difficulty to
derive such an extremely small dark energy scale.

On the other hand, there are some cosmological numerology from a
theoretical speculation related with the observed dark energy. The
dark energy scale coincides with the following relation between the
present Hubble parameter and the Planck scale,
\begin{equation}
{ \rho  }_{ \rm dark }\sim H^{2}_{0}M_{\rm Pl}^{2}\sim 10^{-47}\ {\rm GeV^{4}},
\end{equation}
where $ H_{0}\sim 10^{-42}\ {\rm GeV}$ and the Planck scale is adopted
to be $M_{\rm Pl}\sim 10^{19}\ {\rm GeV}$.  This numerology may
suggest that the current vacuum energy can closely relate with the
current energy scale of the expanding Universe, i.e., the Hubble scale
$H_{0}$ and the Planck-scale physics.  Therefore, there is worth
considering whether the currently-expanding cosmological background
could affect the vacuum energy density through quantum effects.

Theoretically, the dynamical vacuum energy density depends on the
cosmological background, i.e values of the Hubble scale $H$, which has
been indicated in the context of the running vacuum energy scenario.
The scenario is formulated in terms of the renormalization-group (RG)
running cosmological
constant~\cite{Shapiro:2000dz,Sola:2007sv,Shapiro:2008yu,Shapiro:2009dh,Sola:2011qr,Sola:2013gha,Shapiro:2003kv,Shapiro:2003ui,EspanaBonet:2003vk,Shapiro:2004is,Sola:2005et,Sola:2005nh,Babic:2001vv,Babic:2004ev,Markkanen:2014poa},
which actually depends on the renormalization scale
\footnote{ In a low-energy scale ($\mu \ll m$), the one-loop
  $\beta$-function of the cosmological constant
  ${ \beta }_{{ \rho }_{ \Lambda } }$, which includes decoupling effects, 
  is expressed  to be~\cite{Babic:2001vv}
\begin{equation*}
\frac { d{ \rho  }_{ \Lambda  }  }{ d\ln { { \mu } } } =
{ \beta  }_{{ \rho  }_{ \Lambda  } }\simeq \frac { { m }^{ 2 } \mu^{2}}{ { \left( 4\pi  \right)  }^{ 2 } }+\cdots.
\end{equation*}
},
and the physical vacuum energy density can run on the cosmological scale as follows,
\begin{equation}
\frac { d{ \rho  }_{ \Lambda  } }{ d\ln { {\mu} }  } =\frac { 1 }{ { \left( 4\pi  \right)  }^{ 2 } } 
\sum _{ i }^{  }{ \left[ { a }_{ i }{ M }_{ i }^{ 2 }{ \mu }^{ 2 }+{ b }_{ i }{ \mu }^{ 4 }+\O \left( \frac { { \mu }^{ 6 } }
{ { M }^{2}_{ i } }  \right)  \right]  } ,
\end{equation}
where ${ a }_{ i }$, ${ b }_{ i }$ are dimensionless parameters and
$M_{i}$ expresses the mass of the massive field $i$ defined by the UV
theory.  The above equation describes the energy dependence of the
dynamical cosmological constant $ { \rho }_{ \Lambda }$ if we take the
renormalization scale to be $\mu = H\left(t \right) $ with the Hubble
parameter $H\left(t \right)$.  Although this mechanism seems to
realize the cosmological numerology with the Planckian massive fields
$M_{i} \simeq M_{\rm Pl}$ at present, that does not mean that we could
solve the dark energy problem. Still we must put an extra constant term or linear
terms in the above equation by hand~\cite{2012arXiv1204.1864C} because
$ { \rho }_{ \Lambda }$ scales as $\propto H^2$ and is
time-dependent. However it is important that the running vacuum energy
can surely contribute to the energy density even subdominantly~\footnote{
The authors of Refs.~\cite{Sola:2014tta,Gomez-Valent:2014rxa,Sola:2015rra,Sola:2015wwa,Basilakos:2015vra,
Sola:2016vis,Sola:2016jky,Fritzsch:2016ewd,Sola:2016ecz,Sola:2016hnq,Sola:2016lle}
  have reported that the running vacuum models (RVM's) can fit observations
  better than the standard $\Lambda$CDM model with a constant vacuum
  energy $\Lambda$ by adopting SNIa + BAO + $H(z)$ + LSS + BBN + CMB.}.

Theoretically, physical meanings of the RG running cosmological constant in the
context of QFT are unclear, and in addition there are some criticisms about the physical interpretation 
(for the details, see e.g. Refs.~\cite{Foot:2007wn,Shapiro:2008yu,Shapiro:2009dh,Sola:2011qr,Sola:2013gha,Ward:2009wq,Ward:2014sla,Hamber:2013rb}).
For instance, the effective potential is formally scale-invariant
under renormalization group. Therefore, the minimum of the effective
potential, i.e, the renormalized vacuum energy density does not
depend on the renormalization scale $\mu$.  
However, the renormalization-scale invariance of the effective potential merely
comes form a theoretical definition of the problem, and therefore,
we can expect that the RG running effects of the cosmological constant can contribute physically to the
Universe in the same way as the running of the gauge coupling
constants in QED and in QCD~\cite{Mele:2006ji}.  
However, even if the vacuum energy actually run with the cosmological energy-scale,
there still remain some other issues, e.g, why we can take the current
Hubble scale $H_{0}$ as a scale of the running $\mu$ instead of any
other physical scales.  It is still unclear whether the running vacuum
energy appears in the current Universe under the context of the renormalization group.

Therefore, in this paper we revisit the RG running effects of the cosmological constant 
and investigate carefully vacuum contributions of massive quantum fields
by considering the renormalized vacuum energy density on the cosmological background.
By adopting methods discussed in Bunch, Birrell and Davies~\cite{bunch1978quantum,birrell1978application,Bunch:1980vc}, 
we demonstrate that the vacuum energy density of ${ \rho }_{ \rm vacuum } \simeq m^{2}H^{2}$ 
appear as quantum particle production effects of the curved background rather than the RG running of the cosmological constant
\footnote{ So far similar works have been reported to imply
 that the dynamical vacuum energy becomes ${ \rho }_{ \rm vacuum }\simeq m^{2}H^{2}$ in expanding cosmological
 backgrounds. For the details, see e.g. Refs.~\cite{Asorey:2012xq,Maggiore:2010wr,Bilic:2011zm,Bilic:2011rj,
 Hollenstein:2011cz,Hack:2013uyu,Fredenhagen:2013vxa,Padmanabhan:2004qc} and references therein. }
and corresponds to the vacuum field fluctuations
${ \left< { \delta \phi }^{ 2 } \right> } $ growing in proportion to
the Hubble scale $H$.
In the matter dominated Universe, we can obtain the renormalized vacuum energy density as follows:
\begin{eqnarray}
\begin{split}
\rho_{\rm vacuum }
=\ &\frac { { m }^{ 4 }}{ 64{ \pi  }^{ 2 } }
\left(\ln { \frac { { m  }^{ 2 } }{ { \mu }^{ 2 } }  }-\frac{3}{2}\right)  +\frac { 3{ m }^{ 2 }}{ 8{ \pi  }^{ 2 }} \left( \xi -\frac { 1 }{ 6 }  \right) H^{2}
\left(\ln { \frac { { m  }^{ 2 } }{ { \mu }^{ 2 } }  }-\frac{1}{2}\right)+\frac { { m }^{ 2 }{ H }^{ 2 } }{ 96{ \pi  }^{ 2 } }  \\&
-\frac { 81 }{ 64{ \pi  }^{ 2 }} { \left( \xi -\frac { 1 }{ 6 }  \right)  }^{ 2 }H^{4}
\left(\ln { \frac { { m  }^{ 2 } }{ { \mu }^{ 2 } }  }\right)
+\frac { H^{4} }{ 768{ \pi  }^{ 2 }} -\frac { 9 }{ 64{ \pi  }^{ 2 } } \left( \xi -\frac { 1 }{ 6 }  \right)H^{4}+\cdots ,
\end{split}
\end{eqnarray}
where $m$ is the mass of the scalar fields and $\xi$ is the
non-minimal curvature coupling.  Thus the dynamical vacuum energy
appears in the framework of QFT in curved spacetime and provides
outstanding phenomenological contributions to the current Universe.

The present paper is organized as follows. In
Section~\ref{sec:renormalization} we review the renormalized vacuum
energy density and the RG running cosmological constant in Minkowski
spacetime.  Then, we discuss some conceptual and technical
difficulties of the renormalized vacuum energy density.  In
Section~\ref{sec:dynamical} we consider the renormalized vacuum
fluctuations and the renormalized vacuum energy density in curved
spacetime by using the adiabatic regularization and demonstrate that
the curved background can generate the dynamical vacuum energy
density.  In Section~\ref{sec:significance} we discuss the physical
interpretation of the RG running effects of the cosmological constant in Minkowski spacetime and curved spacetime.  Finally, in
Section~\ref{sec:conclusion} we draw the conclusion of our work.

\section{Renormalized vacuum energy density and RG running cosmological constant}
\label{sec:renormalization}
In this section, we discuss a relation of the renormalized vacuum energy density and 
the RG running cosmological constant. 
As an important nature, the renormalized vacuum energy density and 
the renormalized effective potential never depend on the renormalization scale in the framework of QFT.
The physical meanings of the running cosmological constant is hidden by the definition of the renormalization.
However, it does not mean that the vacuum energy does not depend on the cosmological background energy scale.
Following the comprehensive discussion of Ref.\cite{Shapiro:2008yu,Shapiro:2009dh,Sola:2011qr,Sola:2013gha},
we review some conceptual difficulties of the renormalized vacuum energy density.

First of all, let us briefly discuss the gravitational action in the general relativity.
The Einstein-Hilbert action is the well-know and simplest action that yields Einstein's equations,
but, more complicated action involving higher order derivatives is possible and required to the renormalization on the curved spacetime.
The total gravitational action can be defined by 
\begin{equation}
S_{\rm total} \equiv  -\frac { 1 }{ 16\pi { G }_{ N } } \int { { d }^{ 4 }x\sqrt { -g } \left( R+2\Lambda  \right)  }+{ S }_{ \rm HG } +S_{\rm matter},
\end{equation}
where $G_{N}$ is the Newton's gravitational constant and $\Lambda$ is the bare cosmological constant.
The first action express the standard Einstein-Hilbert action with the cosmological constant $\Lambda$,
${ S }_{ \rm HG }$ is the high-order derivative gravitational action and $S_{\rm matter}$ is the matter action.
The high-order gravitational action ${ S }_{ \rm HG }$, which is required to have a renormalizable theory in curved spacetime,
can be written by
\begin{equation}
S_{\rm HG}\equiv  -\int { { d }^{ 4 }x\sqrt { -g } \left( { a }_{ 1 }{ R }^{ 2 }+{ a }_{ 2 }{ C }^{ 2 }
+{ a }_{ 3 }E+{ a }_{ 4 }\Box R  \right)  },
\end{equation}
where $a_{1},a_{2},a_{3},a_{4}$ are high-order derivative gravitational couplings, ${ C }^{ 2 }={ R }_{ \mu \nu \rho \sigma  }{ R }^{ \mu \nu \rho \sigma  }-2{ R }_{ \mu \nu  }{ R }^{ \mu \nu  }+\left( 1/3 \right) { R }^{ 2 }$ is the square of the Weyl tensor and
$E={ R }_{ \mu \nu \rho \sigma  }{ R }^{ \mu \nu \rho \sigma  }-4{ R }_{ \mu \nu  }{ R }^{ \mu \nu  }+{ R }^{ 2 }$
is the Gauss-Bonnet term. The principle of least action with respect to these gravitational actions yields
general Einstein's equations as follows~\cite{Ford:1997hb}
\begin{equation}
\frac { 1 }{ 8\pi { G }_{ N } } { G }_{ \mu \nu  }+{ \rho  }_{ \Lambda  } { g }_{ \mu \nu  }+{ a }_{ 1 }{ H }_{ \mu \nu  }^{ \left( 1 \right)  }
{ +a }_{ 2 }{ H }_{ \mu \nu  }^{ \left( 2 \right)  }{ +a }_{ 3 }{ H }_{ \mu \nu  }={ T }_{ \mu \nu  } ,
\end{equation}
where ${ \rho  }_{ \Lambda  }= \Lambda /{8\pi { G }_{ N }}$ is defined by the cosmological constant,
${ G }_{ \mu \nu  }={ R }_{ \mu \nu  }-\frac { 1 }{ 2 }R{ g }_{ \mu \nu  }$ is the Einstein tensor, 
${ H }_{ \mu \nu  }^{ \left( 1 \right)  }$, ${ H }_{ \mu \nu  }^{ \left( 2 \right)  }$
or ${ H }_{ \mu \nu  }$ are tensors including the high-order derivative terms $R^{2}$, $R_{\mu\nu}R^{\mu\nu}$ or
${ R }_{ \mu \nu \rho \sigma  }{ R }^{ \mu \nu \rho \sigma  }$, and ${ T }_{ \mu \nu  }$ is the energy momentum tensor,
which is defined by
\begin{equation}
{ T }_{ \mu \nu  }=-\frac { 2 }{ \sqrt { -g }  } \frac { \delta S_{\rm matter} }{ \delta { g }^{ \mu \nu  } }.
\end{equation}
The higher derivative terms (${ H }_{ \mu \nu  }^{ \left( 1 \right)  }$, ${ H }_{ \mu \nu  }^{ \left( 2 \right)  }$
and ${ H }_{ \mu \nu  }$) express short distance effects and 
do not contribute to  Einstein's equations at long distances or weak gravitational fields.
Therefore, standard Einstein's equations with the cosmological constant $\Lambda$, which describes 
the gravitational dynamics of the cosmological scale, is formally given by 
\begin{equation}
{ R }_{ \mu \nu  }-\frac { 1 }{ 2 }R{ g }_{ \mu \nu  }+\Lambda{ g }_{ \mu \nu  } =8\pi G_{N}{ T }_{ \mu \nu  },
\end{equation}
where ${ R }_{ \mu \nu  }$ is the Riemann tensor, $R$ is the Ricci scalar, ${ g }_{ \mu \nu  }$ is the metric of spacetime.
For simplicity, we assume the following matter action with the scalar field $\phi$ as
\begin{equation}
S_{\rm matter} =  \int { { d }^{ 4 }x\sqrt { -g }\mathcal{L}_{\phi} }=
\int { { d }^{ 4 }x\sqrt { -g } \left[ \frac { 1 }{ 2 } { g }^{ \mu \nu  }{ \partial  }_{ \mu  }\phi { \partial  }_{ \nu  }\phi 
- V\left( \phi  \right)  \right]  } ,
\end{equation}
where $\mathcal{L}_{\phi}$ is the matter Lagrangian for the classical
scalar field $\phi$ and therefore, the scalar potential
$V\left( \phi \right)$ is the classical potential.  The
energy-momentum tensor derived from this matter action
$S_{\rm matter} $ can be written as follows
\begin{equation}
{ T }_{ \mu \nu  }={ \partial  }_{ \mu  }\phi { \partial  }_{ \nu  }\phi -{ g }_{ \mu \nu  }\left( \frac { 1 }{ 2 }{ g }^{ \rho \sigma  }
 { \partial  }_{ \rho  }\phi { \partial  }_{\sigma  }\phi + V\left( \phi  \right) \right) .
\end{equation}
Here, we rewrite standard Einstein's equations by transposing the cosmological constant term $\Lambda$ into 
the energy momentum tensor ${ T }_{ \mu \nu  }$ as 
\begin{equation}
{ R }_{ \mu \nu  }-\frac { 1 }{ 2 }R{ g }_{ \mu \nu  }=8\pi G_{N}{ T }_{ \mu \nu  }^{\Lambda}, \quad
{ T }_{ \mu \nu  }^{\Lambda} ={ \rho  }_{ \Lambda  }  +{ T }_{ \mu \nu  }.
\end{equation}
Here, we divide the energy momentum tensor into the vacuum part and the matter part 
in the context of QFT as follows:
\begin{equation}
{ R }_{ \mu \nu  }-\frac { 1 }{ 2 }R{ g }_{ \mu \nu  }=8\pi G_{N}\left(  \left<{ T }_{ \mu \nu  }^{\Lambda} \right>  
+{ T }_{ \mu \nu  }^{\Lambda}  \right),
\end{equation}
with
\begin{equation}
 \left<{ T }_{ \mu \nu  }^{\Lambda}  \right> \equiv \left< { 0 }|{ { T }_{ \mu \nu  }^{\Lambda} }|{ 0 } \right>  
\equiv{ g }_{ \mu \nu  }{ \rho  }_{\rm vacuum  },
\end{equation}
where $\left< { T }_{ \mu \nu  }^{\Lambda} \right> $ expresses the ground value of 
the energy-momentum tensor ${ T }_{ \mu \nu  }^{\Lambda} $ and
corresponds to the vacuum energy density ${ \rho  }_{ \rm vacuum  } $.
The vacuum energy density ${ \rho  }_{ \rm vacuum  } $ originates from
the cosmological constant ${ \rho  }_{ \Lambda  }$ of Einstein's equations,
and the classical vacuum energy via the spontaneous symmetry breaking (SSB) 
or the quantum corrections of vacuum field fluctuations. 
Although there are several sources of the vacuum energy density ${ \rho  }_{ \rm vacuum  } $,
the quantum corrections of the massive fields would bring most of the large contribution to
the vacuum energy density.

Formally, the quantum corrections of the vacuum energy density are consistent with the quantum zero-point energy density
and given as the following equation
\begin{eqnarray}
\rho_{\rm zero}
=\frac{1}{2}\int { \frac { { d }^{ 3 }k }{ { \left( 2\pi  \right)  }^{ 3 } }
 \sqrt { { k }^{ 2 }+{ m }^{ 2 } }  },
\end{eqnarray}
where the zero-point vacuum energy is formally infinite and needs the renormalization
in order to obtain the physical finite result.
The divergences of the zero-point vacuum energy are usually neglected by the normal ordering of 
the operators or removed by the bare cosmological constant as we will see later.
If we adopt the dimensional regularization,
the divergences of the zero-point vacuum energy are regularized as the following equation
\footnote{
By adopting the momentum cut-off $\Lambda_{\rm UV}$,
the divergences of the zero-point vacuum energy are regularized as follows
\begin{eqnarray}
\rho_{\rm zero }
&=&\frac{1}{2}\int^{\Lambda_{\rm UV}} { \frac { { d }^{ 3 }k }{ { \left( 2\pi  \right)  }^{ 3 } }
 \sqrt { { k }^{ 2 }+{ m }^{ 2 } }  } \\
 &=&\frac { { \Lambda_{\rm UV}^{ 4 }   }}{ 16{ \pi  }^{ 2 } } +\frac { { { m }^{ 2 }\Lambda_{\rm UV}^{ 2 }  } }{ 16{ \pi  }^{ 2 } } +\frac { { m }^{ 4 } }{ 64{ \pi  }^{ 2 } } \log { \left( \frac { { m }^{ 2 } }{ { \Lambda_{\rm UV}^{ 2 }   }}  \right)  } +\cdots ,
\end{eqnarray}
where quartic or quadratic divergences in this expression are unphysical and removed by the renormalization.
When the cut-off $\Lambda_{\rm UV}$ corresponds to the Planck scale $M_{\rm Pl}$,
the first term expresses the fact that 
the vacuum energy density ${ \rho  }_{\rm vacuum}$ has to be fine-tuned to 123 orders of magnitude
in order to accommodate the observed dark energy.}
\begin{equation}
{ \rho  }_{ \rm zero}=\frac { { m }^{ 4 } }{ 64{ \pi  }^{ 2 } } \left[ \ln { \left( \frac { { m }^{ 4 } }{ { \mu  }^{ 2 } }  \right)  } 
-\frac { 1 }{ \epsilon  } -\log { 4{ \pi  } } +\gamma -\frac { 3 }{ 2 }  \right] ,
\end{equation}
where $\mu$ is the subtraction scale of dimensional regularization,
$\epsilon$ is the regularization parameter and $\gamma$ is the Euler's constant.

In the standard QFT, the divergence terms are absorbed by the bare cosmological constant term 
${ \rho  }_{ \Lambda  }$ of the Einstein-Hilbert action.
We split the bare term ${ \rho  }_{ \Lambda  }$ as ${ \rho  }_{ \Lambda  } ={ \rho  }_{ \Lambda  }\left(\mu\right)
+\delta{ \rho  }_{ \Lambda  }$
where the counterterm $\delta{ \rho  }_{ \Lambda  }$ depends on the regularization and renormalization scheme.
The counterterm in $\overline {\rm MS }$ scheme is given by
\begin{equation}
\delta{ \rho  }_{ \Lambda  }=\frac { { m }^{ 4 } }{ 4{ \left( 4\pi  \right)  }^{ 2 } } \left(
\frac { 1 }{ \epsilon  } +\log { 4{ \pi  } } -\gamma  \right).
\end{equation}
By absorbing the divergence terms into the counterterm $\delta{ \rho  }_{ \Lambda  }$, 
we can obtain the renormalized zero-point energy density as follows
\begin{equation}
{ \rho  }_{\rm zero}\left(\mu\right)={ \rho  }_{ \rm zero}+\delta{ \rho  }_{ \Lambda  }
=\frac { { m }^{ 4 }}{ 64{ \pi  }^{ 2 } }
 \left(  \ln {\frac { { m }^{ 4 }  }{ { \mu  }^{ 2 } }  }-\frac{3}{2} \right) ,
\end{equation}
where we obtain the finite expression without UV divergences.
Thus, the renormalized vacuum energy density at one-loop order is given by
\begin{equation}
{ \rho  }_{\rm vacuum}={ \rho  }_{ \Lambda  }\left(\mu\right)+
{ \rho  }_{\rm zero}\left(\mu\right)={ \rho  }_{ \Lambda  }\left(\mu\right)+\frac { { m }^{ 4 }}{ 64{ \pi  }^{ 2 } }
 \left(  \ln {\frac { { m }^{ 4 } }{ { \mu  }^{ 2 } }  }-\frac{3}{2} \right) \label{eq:dsldkfgsdg},
\end{equation}
where the interpretation of the renormalization-scale $\mu$ raises some problems.
The renormalization scale $\mu$, which is introduced in the dimensional regularization, should be exchanged for 
some physical quantity, e.g. the energy of scattering of particles, the size of the finite system, 
the VEV of the scalar field or the temperature of the system.
In cosmological situations, it is reasonable to take the current Hubble scale $H_{0}$ as the renormalization-scale $\mu$
against the observed dark energy as ${ \rho  }_{\rm vacuum}\left(\mu \simeq H_{0}\right) \simeq { \rho  }_{\rm dark}$.
However, there is no rigid reason that to take the current Hubble scale $H_{0}$ as the renormalization-scale $\mu$
instead of the current temperature $T_{0}$ or the electroweak breaking scale $M_{\rm EW}$.
The physical interpretation of the renormalization-scale $\mu$ is still unclear in
the renormalized vacuum energy density.

From the standard procedure of QFT, we require the cancellation of 
$\mu$-dependence for the renormalized vacuum energy density as
\begin{equation}
\frac { d{ \rho  }_{\rm vacuum}}{ d\ln { { \mu } }  } =0 .
\end{equation}
Thus, we can estimate the one-loop $\beta$-function of the cosmological constant as follows
\begin{equation}
\frac { d{ \rho  }_{ \Lambda  } }{ d\ln { { \mu } }  } =
{ \beta  }_{{ \rho  }_{ \Lambda  } }=\frac { { m }^{ 4 } }{ 2{ \left( 4\pi  \right)  }^{ 2 } } ,
\end{equation}
which governs the renormalization-group running of the cosmological constant.
The physical renormalized vacuum energy density dressed in quantum fluctuations is 
only determined by the experimental observation at the energy scale
\footnote{
The fine-tuning problem of the renormalized vacuum energy density still remains as
\begin{equation}
{ \rho  }_{\rm  vacuum }=O\left( m^{4} \right) -O\left( m^{4} \right) ,
\end{equation}
where we assume the renormalization-scale as $\mu \gtrsim m$. The finite naturalness 
is somewhat alleviated in comparison with the naive cut-off $\Lambda_{\rm UV}$.
However, if we consider the Planck scale or GUT scale massive fields, 
the renormalized vacuum energy density ${ \rho  }_{\rm vacuum}$ 
suffers from the fine-tuning of 123 orders, and therefore, 
the situation does not improve. However, when we consider the low-energy scale as $\mu \ll m$,
the fine-tuning of the renormalized vacuum energy density is 
well alleviated as the following
\begin{equation}
{ \rho  }_{\rm  vacuum }=O\left( m^{2}\mu^{2} \right) -O\left( m^{2}\mu^{2} \right) ,
\end{equation}
where the decoupling effects suppress the quantum corrections. If we set $\mu \simeq H_{0}$,
there is no need to fine-tune the renormalized vacuum energy density ${ \rho  }_{\rm vacuum}$ unless
trans-Planckian massive fields as $m^{2} \gtrsim  M_{\rm Pl}$ exist.
However, there is no rigid reason that to assume the current Hubble scale $H_{0}$
as the renormalization-scale $\mu$. Furthermore, 
the current status of the LHC experiments has aggravated the Higgs naturalness
as the fine-tuning of the vacuum energy 
(see e.g. Ref.\cite{Giudice:2013nak,Farina:2013mla,Dine:2015xga,Matsui:2016cls} for the details).}.
Although the cosmological constant run under the renormalization-group, formally, 
the renormalized vacuum energy does not run with the renormalization-scale $\mu$.
The renormalization-scale independence of the observable quantities 
is the main feature of the renormalization group. 
However, why the running of the cosmological constant doesn't seem to provide 
physical contributions in contrast to the running of the gauge coupling.
Unfortunately, we can not say anything from the renormalized vacuum energy density of Eq.~(\ref{eq:dsldkfgsdg}).
For these issues we discuss more specifically in Section~\ref{sec:significance}.

In the dynamical background, e.g, the curved background or the vacuum condensate background, however, 
the vacuum energy density grows in proportion to the background energy scale as 
${ \rho  }_{\rm vacuum}=\mathcal{O}\left(H^{4}\right), \mathcal{O}\left(\phi^{4}\right)$,
and therefore, the renormalization scale $\mu$ of the renormalized vacuum energy density would
correspond to the physical energy scale rather than the arbitrary parameter.
Thus, we can estimate the physical vacuum energy from the RG running of the cosmological constant 
if we take the renormalization-scale $\mu$ as the energy-scale of the dynamical background.
Here, we rewrite the renormalized vacuum energy density of Eq.~(\ref{eq:dsldkfgsdg})
\begin{equation}
{ \rho  }_{\rm vacuum}={ \rho  }_{ \Lambda  }+\frac { { m }^{ 4 }}{ 64{ \pi  }^{ 2 } }
 \left(  \ln {\frac { { m }^{ 4 } }{ { \mu  }^{ 2 } }  }-\frac{3}{2} \right) ,
\end{equation}
where the mass $m$ or the cosmological constant $\rho_{\Lambda }$ are only determined by $\mu$.
However, in the low-energy scale $\mu \ll m$, the above expression is not correct due to the decoupling effects. 
The correct expression including the decoupling effects is written by
\begin{equation}
{ \rho  }_{\rm vacuum}=\mu^{4}+\frac { { m }^{ 2}\mu^{2}}{ 64{ \pi  }^{ 2 } }+\cdots .
\end{equation}
By taking the renormalization scale as $\mu \simeq H$, we can obtain 
the dynamical vacuum energy in the cosmological background as follows
\begin{equation}
{ \rho  }_{\rm vacuum}\simeq H^{4}+\frac { { m }^{ 2}H^{2}}{ 64{ \pi  }^{ 2 } }+\cdots .
\end{equation}
However, the above discussion is not rigorous proof of the running vacuum energy in the curved background. 
From the viewpoint of the QFT in curved spacetime, rigidly,  
we must calculate the renormalized vacuum energy density by using some technical renormalization methods.
Therefore, let us consider the renormalized vacuum energy density on curved background in the next section.

\section{Renormalized vacuum field fluctuations and renormalized vacuum energy density in curved background}
\label{sec:dynamical}
In previous section, we have discussed the ambiguity of the renormalized vacuum energy density and the running cosmological constant.
Although the running effects of the cosmological constant indicate the dynamical vacuum energy 
on the cosmological background, we do not obtain a strict proof of the dynamical vacuum energy.
Thus, in this section, let us consider the renormalized vacuum field fluctuations
and the renormalized vacuum energy density in the framework of the QFT in curved spacetime.

For simplicity, we consider the Friedmann-Lemaitre-Robertson-Walker (FLRW) metric as
$g_{\mu\nu}={\rm diag}\left( -1,{ a }^{ 2 }\left( t \right)  ,{ a }^{ 2 }\left( t \right) { r }^{ 2 },
{ a }^{ 2 }\left( t \right) { r }^{ 2 }\sin^{2} { \theta  }  \right)$ for the spatially flat Universe. 
Therefore, the Ricci scalar can be written as follows
\begin{equation}
R=6\left[ 
{ \left( \frac { \dot { a }  }{ a }  \right)  }^{ 2 }+\left( \frac { \ddot { a }  }{ a }  \right)  
\right]=6\left(\frac{a''}{a^{3}}\right)\label{eq:dgdgfdgedg}, 
\end{equation}
where the conformal time $\eta$ is defined by $d\eta=dt/a$.
In the matter dominated Universe,
the scale factor becomes $a\left(t\right) =t^{2/3}$ and 
the Ricci scalar is estimated to be $R=3H^{2}$.
On the other hand, 
in the de Sitter Universe, the scale factor becomes $a\left(t\right) =e^{Ht}$ or  $a\left(\eta\right) =-1/H\eta$, 
the Ricci scalar is estimated to be $R=12H^{2}$. 
Then, we assume the matter action for the scalar field $\phi$ in curved spacetime as
\begin{equation}
S_{\rm matter} =-\int { { d }^{ 4 }x\sqrt { -g } \left( \frac { 1 }{ 2 } { g }^{ \mu \nu  }
{ \partial   }_{ \mu  }\phi {\partial  }_{ \nu  }\phi +\frac{1}{2}\left(m^{2}+\xi R\right)\phi^{2}  \right)  } \label{eq:dddddddfg}.
\end{equation}
The Klein-Gordon equation is given as the following
\begin{equation}
\Box \phi\left(\eta ,x\right)+\left(m^{2}+\xi R\right)\phi\left(\eta ,x\right)  =0 \label{eq:dsssssdg},
\end{equation}
where $\Box$ express the generally covariant d'Alembertian operator, 
$\Box =g^{\mu\nu}{ \nabla  }_{ \mu  }{  \nabla   }_{ \nu  }=1/\sqrt { -g } { \partial  }_{ \mu  }\left( \sqrt { -g } { \partial  }^{ \mu  } \right) $
and $\xi$ is the non-minimal curvature coupling constant. 
Then, we decompose the scalar field  $\phi \left(\eta ,x\right)$  into the classic field and the quantum field as
\begin{equation}
\phi \left(\eta ,x\right)=\phi \left(\eta \right)+\delta \phi \left(\eta ,x \right) \label{eq:kkkkgedg},
\end{equation}
where we assume the vacuum expectation value of the scalar field is 
$\phi\left(\eta \right)=\left< 0  \right| { \phi \left(\eta ,x\right) }\left| 0 \right>$ and 
$\left< 0  \right| { \delta\phi \left(\eta ,x\right) }\left| 0 \right>=0$.
In the one-loop approximation, we obtain the Klein-Gordon equation as the following
\begin{eqnarray}
&&\Box \phi +\left(m^{2}+\xi R\right)\phi =0 \label{eq:lklkedg},\\
&&\Box \delta \phi +\left(m^{2}+\xi R\right) \delta \phi =0 \label{eq:dppppg}.
\end{eqnarray}
The quantum field $\delta \phi$ is decomposed into each $k$ modes as the following,
\begin{equation}
\delta \phi \left( \eta ,x \right) =\int { { d }^{ 3 }k\left( { a }_{ k }\delta { \phi  }_{ k }\left( \eta ,x \right) +{ a }_{ k }^{ \dagger  }\delta { \phi  }_{ k }^{ * }\left( \eta ,x \right)  \right)  }  \label{eq:ddfkkfledg},
\end{equation}
where we assume 
\begin{equation}
{ \delta \phi  }_{ k }\left( \eta ,x \right) =\frac { { e }^{ ik\cdot x } }{ { \left( 2\pi  \right)  }^{ 3/2 }
\sqrt { C\left( \eta  \right)  }  } \delta { \chi  }_{ k }\left( \eta  \right)  \label{eq:slkdlkgdg},
\end{equation}
with $C\left(\eta \right)=a^{2}\left(\eta \right)$.
Thus, the expectation field value $\left<{  \delta \phi   }^{ 2 } \right>$  can be given by
\begin{eqnarray}
\left< 0  \right| { \delta \phi^{2}  }\left| 0 \right>&=&\int { { d }^{ 3 }k{ \left| \delta { \phi  }_{ k }\left( \eta ,x \right)  \right|  }^{ 2 } } \label{eq:ddkg;ergedg},\\
&=&\frac { 1 }{ 2{ \pi  }^{ 2 }C\left( \eta  \right)  } \int _{ 0 }^{ \infty  } { dk { k }^{ 2 }{ \left| \delta { \chi  }_{ k } \right|  }^{ 2 } }  \label{eq:xkkgdddgedg},
\end{eqnarray}
where $\left<{  \delta \phi   }^{ 2 } \right>$ has quadratic or logarithmic ultraviolet divergences 
and needs to be regularized.
From Eq.~(\ref{eq:dppppg}), the Klein-Gordon equation for the quantum field $\delta \chi$ can be given by
\begin{equation}
{ \delta \chi}''_{ k }+{ \Omega  }_{ k }^{ 2 }\left( \eta  \right) { \delta \chi  }_{ k }=0 \label{eq:dlkrlekg},
\end{equation}
where ${ \Omega  }_{ k }^{ 2 }\left( \eta  \right) ={ \omega  }_{ k }^{ 2 }\left( \eta  \right) + C\left( \eta  \right) \left( \xi -1/6 \right) R $ and ${ \omega  }_{ k }^{ 2 }\left( \eta  \right) ={ k }^{ 2 }+C\left( \eta  \right){ m }^{ 2 } $.
The Wronskian condition is given as
\begin{equation}
{ \delta\chi  }_{ k }{ \delta\chi  }_{ k }^{ * }-{ \delta\chi  }_{ k }^{ * }{ \delta\chi  }_{ k }= i\label{eq:dddrrrg},
\end{equation}
which ensures the canonical commutation relations for the field operator $\delta\chi$ as follows
\begin{eqnarray}
\bigl[ { a }_{ k },{ a }_{ k' } \bigr] &=&\bigl[{ a }_{ k }^{ \dagger  },{ a }_{ k' }^{ \dagger  } \bigr]=0 \label{eq:dfphipedg},\\ 
\bigl[{ a }_{ k },{ a }_{ k' }^{ \dagger  } \bigr] &=&\delta \left( k-k' \right) \label{eq:oeuoegedg}.
\end{eqnarray}
The adiabatic vacuum $\left| 0 \right> $ is defined as the vacuum state which is annihilated by all the operators ${ a }_{ k }$
and set ${ \delta\chi  }_{ k }\left(\eta\right)$ to be a positive-frequency mode.
The adiabatic (WKB) approximation to the mode function ${ \delta\chi  }_{ k }\left(\eta\right)$ is written by 
\begin{equation}
{ \delta\chi  }_{ k }\left(\eta\right)=\frac { 1 }{ \sqrt { 2{ W }_{ k }\left( \eta  \right) C\left( \eta  \right)  }  } 
\exp\left( -i\int { { W }_{ k }\left( \eta  \right) d\eta  }  \right) \label{eq:jgskedg},
\end{equation}
with
\begin{equation}
{ W }_{ k }^{ 2 }={ \Omega  }_{ k }^{ 2 }-\frac { 1 }{ 2 } \frac { { W }''_{ k } }{ { W }_{ k } } 
+\frac { 3 }{ 4 } \frac { { \left( { W }'_{ k } \right)  }^{ 2 } }{ { W }_{ k }^{ 2 } } \label{eq:sklsdkldg}.
\end{equation}
We can obtain the WKB solution by solving iteratively Eq.~(\ref{eq:sklsdkldg})
and the lowest-order WKB solution ${ W }^{0}_{ k }$ is written by
\begin{equation}
\left({ W }^{0}_{ k }\right)^{2}={ \Omega  }_{ k }^{2}\label{eq:flkldg}.
\end{equation}
The first-order WKB solution ${ W }^{1}_{ k }$ is written by
\begin{equation}
\left({ W }^{1}_{ k }\right)^{2}={ \Omega  }_{ k }^{2}-\frac { 1 }{ 2 } \frac { \left({ W}^{0}_{ k }\right)'' }{{ W }^{0}_{ k } } 
+\frac { 3 }{ 4 } \frac {\left({{ W}^{0}_{ k }}'\right)^{2}  }{ \left({ W}^{0}_{ k }\right)^{2} } \label{eq:skdl;lgedg}.
\end{equation}
For the high-order WKB solution, we can obtain the following expression 
\begin{eqnarray}
{ W }_{ k }&\simeq&{ \omega  }_{ k }+\frac { 3\left( \xi -1/6 \right)  }{ 4{ \omega  }_{ k } } \left( 2D'+{ D }^{ 2 } \right) -\frac { { m }^{ 2 }C }{ 8{ \omega  }_{ k }^{ 3 } } \left( D'+{ D }^{ 2 } \right) +\frac { 5{ m }^{ 4 }{ C }^{ 2 }{ D }^{ 2 } }{ 32{ \omega  }^{ 5 } } \nonumber \\
&&+\frac { { m }^{ 2 }C }{ 32{ \omega  }_{ k }^{ 5 } } \left( D'''+4D'D+3{ D' }^{ 2 }+6D'D^{ 2 }+D^{ 4 } \right) \nonumber \\
&&-\frac { { m }^{ 4 }{ C }^{ 2 } }{ 128{ \omega  }_{ k }^{ 7 } } \left( 28D''D+19{ D' }^{ 2 }+122{ D' }^{ 2 }+47D^{ 4 } \right)  \nonumber \\ 
&&+\frac { { 221m }^{ 6 }{ C }^{ 3 } }{ 256{ \omega  }_{ k }^{ 9 } } \left( D'D^{ 2 }+D^{ 4 } \right) 
-\frac { 1105{ m }^{ 8 }{ C }^{ 4 }{ D }^{ 4 } }{ 2048{ \omega  }_{ k }^{ 11 } }  \nonumber \\
&&-\frac { \left( \xi -1/6 \right)  }{ 8{ \omega  }_{ k }^{ 3 } } \left( 3D'''+3D''D+3{ D' }^{ 2 } \right) \nonumber \\
&&+\left( \xi -1/6 \right) \frac { { m }^{ 2 }C }{ 32{ \omega  }_{ k }^{ 5 } } 
\left( 30D''D+18{ D' }^{ 2 }+57D'D^{ 2 }+9D^{ 4 } \right)  \nonumber \\
&&-\left( \xi -1/6 \right) \frac { { 75m }^{ 4 }{ C }^{ 2 } }{ 128{ \omega  }_{ k }^{ 7 } } \left( 2D'D^{ 2 }+D^{ 4 } \right) \nonumber \\
&&-\frac { { \left( \xi -1/6 \right)  }^{ 2 } }{ 32{ \omega  }_{ k }^{ 3 } } \left( 36{ D' }^{ 2 }+36D'D^{ 2 }+9D^{ 4 } \right)+\cdots \label{eq:oekgkf},\end{eqnarray}
where $D=C'/C$. The vacuum expectation value ${ \left< { \delta \phi   }^{ 2 } \right>  }$  of the lowest-order WKB solution
has clearly divergences as 
\begin{eqnarray}
{ \left< { \delta \phi   }^{ 2 } \right>  }_{ \rm div }
=\frac { 1 }{ 4{ \pi  }^{ 2 }C\left( \eta  \right)  } \int _{ 0 }^{ \infty  }{ \frac { { k }^{ 2 } }{ { \Omega }_{ k } } dk } \label{eq:dlajshegedg}.
\end{eqnarray}
However, the high-order corrections do not have such divergences
and UV finite (see e.g. Ref.\cite{Ringwald:1987ui} for the details).
The adiabatic regularization is not just a mathematical method of regularizing divergent integrals as 
cut-off regularization or dimensional regularization.
The divergences of the original expression ${ \left< { \delta \phi   }^{ 2 } \right>  }$,
which come from the large $k$  modes, are the same as the divergences of the lowest-order WKB solution
${ \left< { \delta \phi   }^{ 2 } \right>_{\rm div}   }$.
Thus, by subtracting the lowest-order WKB solution ${ \left< { \delta \phi   }^{ 2 } \right>_{\rm div}   }$
from the original expression ${ \left< { \delta \phi   }^{ 2 } \right>  }$, we can obtain the finite vacuum field fluctuation as follows:
\begin{eqnarray}
{ \left< { \delta \phi   }^{ 2 } \right>  }_{ \rm ren }
&=&{ \left< {  \delta \phi   }^{ 2 } \right>  }-{ \left< {  \delta \phi   }^{ 2 } \right>  }_{ \rm div },\\
&=&\frac { 1 }{ 4{ \pi  }^{ 2 }C\left( \eta  \right)  }  
\Biggl[ \int _{ 0 }^{ \infty  }{ 2k^{2}{ \left| \delta { \chi  }_{ k } \right|  }^{ 2 }dk } 
-\int _{ 0 }^{ \infty  }{ \frac { { k }^{ 2 } }{ { \Omega  }_{ k } } dk } \Biggr] \label{eq:dkhkkegedg},
 \end{eqnarray}
where the renormalized vacuum filed fluctuations include the dynamical particle production effects
in curved spacetime. 
The adiabatic regularization~\cite{birrell1978application,
Bunch:1980vc,Fulling:1974pu,Fulling:1974zr,Parker:1974qw,Anderson:1987yt,
Paz:1988mt,Haro:2010zz,Haro:2010mx}  is the extremely powerful method to remove the ultraviolet divergences
of the vacuum field fluctuations or the vacuum energy density,
and has been shown to be equivalent to 
the point-splitting regularization in Ref.\cite{birrell1978application,Anderson:1987yt}.

For simplicity, we consider the vacuum field fluctuations in the massive conformally coupled case ($m \gg H$ and $\xi = 1/6$). 
In this case, the adiabatic (WKB) conditions (${ \Omega  }_{ k }^{ 2 }>0$ and 
$\left| { \Omega ' }_{ k }/{ \Omega  }_{ k }^{ 2 } \right| \ll 1$) are satisfied and
we can obtain the following adiabatic (WKB) solution as
\begin{eqnarray}
{ W }_{ k }&=&{ \omega  }_{ k } -\frac { { m }^{ 2 }C }{ 8{ \omega  }_{ k }^{ 3 } } \left( D'+{ D }^{ 2 } \right)
 +\frac { 5{ m }^{ 4 }{ C }^{ 2 }{ D }^{ 2 } }{ 32{ \omega  }^{ 5 } }+\cdots ,\\
&=&{ \omega  }_{ k }-\frac { 1 }{ 8 } \frac { { m }^{ 2 }{ C }'' }{ { \omega  }_{ k }^{ 3 } } 
+\frac { 5 }{ 32 } \frac { { { m }^{ 4 }\left( C' \right)  }^{ 2 } }{ { \omega  }_{ k }^{ 5 } }+\cdots \label{eq:ofhfhgedg}.
\end{eqnarray}
Therefore, we  obtain the following expression
\begin{equation}
\frac{1}{{ W }_{ k }}\simeq\frac{1}{{ \omega  }_{ k }}
+\frac { 1 }{ 8 } \frac { { m }^{ 2 }{ C }'' }{ { \omega  }_{ k }^{ 5 } } 
-\frac { 5 }{ 32 } \frac { { { m }^{ 4 }\left( C' \right)  }^{ 2 } }{ { \omega  }_{ k }^{ 7 } }+\cdots \label{eq:fhfhgedg}.
\end{equation}
The renormalized vacuum field fluctuations from Eq.~(\ref{eq:dkhkkegedg}) can be given as follows:
\begin{eqnarray}
{ \left< { \delta \phi   }^{ 2 } \right>  }_{ \rm ren }
&=&\lim _{ \Lambda_{\rm UV}  \rightarrow  \infty   }\frac { 1 }{ 4{ \pi  }^{ 2 }C\left( \eta  \right)  }  
\Biggl[ \int _{ 0 }^{ \Lambda_{\rm UV}   }{ 2k^{2}{ \left| \delta { \chi  }_{ k } \right|  }^{ 2 }dk } 
-\int _{ 0 }^{ \Lambda_{\rm UV}   }{ \frac { { k }^{ 2 } }{ { \Omega}_{ k } } dk } \Biggr] , \\
&\simeq&\lim _{ \Lambda_{\rm UV}  \rightarrow  \infty   }\frac { 1 }{ 4{ \pi  }^{ 2 }C\left( \eta  \right)  }  
\Biggl[ \int _{ 0 }^{ \Lambda_{\rm UV}   }{ \frac { { k }^{ 2 } }{ {W}_{ k } }dk } 
-\int _{ 0 }^{ \Lambda_{\rm UV}   }{ \frac { { k }^{ 2 } }{ { \omega}_{ k } } dk } \Biggr] , \\
&\simeq&\lim _{ \Lambda_{\rm UV}  \rightarrow  \infty   }\frac { 1 }{ 4{ \pi  }^{ 2 }C\left( \eta  \right)  }
 \Biggl[ \int _{0}^{ \Lambda_{\rm UV}   }{ \frac { { k }^{ 2 } }{ { \omega  }_{ k } } dk} 
 -\int _{0 }^{ \Lambda_{\rm UV}   }{ \frac { { k }^{ 2 } }{ { \omega  }_{ k } } dk} \nonumber \\
&&+\frac { { m }^{ 2 }C'' }{ 8 } \int _{ 0 }^{ \Lambda_{\rm UV}   }{ \frac { { k }^{ 2 } }{ { \omega  }_{ k }^{ 5 } } dk
+\frac { { 5m }^{ 4 }{ \left( C' \right)  }^{ 2 } }{ 32 } \int _{ 0}^{ \Lambda_{\rm UV}   }
 { \frac { { k }^{ 2 } }{ { \omega  }_{ k }^{ 7 } } dk }  }+\cdots\Biggr]
\label{eq:fhgegedg}.
\end{eqnarray}
 Therefore, we can obtain the following expression 
\begin{equation}
{ \left< { \delta \phi   }^{ 2 } \right>  }_{ \rm ren }
=\lim _{ \Lambda_{\rm UV} \rightarrow  \infty   }\frac { 1 }{ 4{ \pi  }^{ 2 }C\left( \eta  \right)  }  
\Biggl[ \frac { { m }^{ 2 }C'' }{ 8 } \int _{ 0 }^{ \Lambda_{\rm UV}   }{ \frac { { k }^{ 2 } }{ { \omega  }_{ k }^{ 5 } } dk
+\frac { { 5m }^{ 4 }{ \left( C' \right)  }^{ 2 } }{ 32 } \int _{ 0}^{ \Lambda_{\rm UV}   }
 { \frac { { k }^{ 2 } }{ { \omega  }_{ k }^{ 7 } } dk }  } +\cdots \Biggr]\label{eq:oehf:lhg}.
 \end{equation}
By using the UV cut-off $\Lambda_{\rm UV} $, we can simply estimate the dominated terms 
of the renormalized vacuum field fluctuation as follows:
\begin{eqnarray}
&&\lim _{ \Lambda_{\rm UV}  \rightarrow  \infty   }\frac { 1 }{ 4{ \pi  }^{ 2 }C\left( \eta  \right)  }  
\Biggl[ \frac { { m }^{ 2 }C'' }{ 8 } \int _{ 0 }^{ \Lambda_{\rm UV}   }{ \frac { { k }^{ 2 } }{ { \omega  }_{ k }^{ 5 } } dk
-\frac { { 5m }^{ 4 }{ \left( C' \right)  }^{ 2 } }{ 32 } \int _{ 0 }^{ \Lambda_{\rm UV}   }{ \frac { { k }^{ 2 } }{ { \omega  }_{ k }^{ 7 } } dk }  }\Biggr]
\nonumber \\
=&&\lim _{ \Lambda_{\rm UV}  \rightarrow  \infty   }\frac { 1 }{ 4{ \pi  }^{ 2 }C\left( \eta  \right)  }  \Biggl[ \frac { { m }^{ 2 }C'' }{ 8 } \frac { { \Lambda_{\rm UV}^{ 3 }   } }{ 3{ m }^{ 2 }C{ \left( { \Lambda_{\rm UV}^{ 2 }   }+{ m }^{ 2 }C \right)  }^{ 3/2 } } 
-\frac { 5{ m }^{ 4 }{ \left( C' \right)  }^{ 2 } }{ 32 } \frac { { { 5m }^{ 2 }C\Lambda_{\rm UV}^{ 3 }   }+2{ \Lambda_{\rm UV}^{ 5 }   } }{ 15{ m }^{ 4 }{ C }^{ 2 }{ \left( { \Lambda_{\rm UV}^{ 2 }   }+{ m }^{ 2 }C \right)  }^{ 5/2 } }   \Biggr], \nonumber \\
=&&-\frac { 1 }{ 96{ \pi  }^{ 2 }C\left( \eta  \right)  } 
\left[ \frac { 1 }{ 2 } { \left( \frac { C' }{ C }  \right)  }^{ 2 }-\frac { C'' }{ C }  \right]=\frac { 1 }{ 48{ \pi  }^{ 2 } } \frac { a'' }{ { a }^{ 3 } } 
=\frac { R }{ 288{ \pi  }^{ 2 } }
\label{eq:fhkgkgedg}.
\end{eqnarray}
Therefore, we have the renormalized vacuum field fluctuations in the massive conformal coupling case as follows:
\begin{eqnarray}
{ \left< { \delta \phi   }^{ 2 } \right>  }_{ \rm ren }
=\frac { R }{ 288{ \pi  }^{ 2 } }+\mathcal{O}\left(R^{2}\right)+\cdots \label{eq:oesdhg}.
\end{eqnarray}
In the matter dominated Universe, the Ricci scalar becomes $R=3H^{2}$ and 
the renormalized vacuum field fluctuations are  ${ \left< { \delta \phi   }^{ 2 } \right>  }_{ \rm ren }\simeq H^{2}/96\pi^{2}$.
On the other hand, in the de Sitter Universe, the Ricci scalar is estimated to be $R=12H^{2}$ and  
the renormalized vacuum field fluctuations become ${ \left< { \delta \phi   }^{ 2 } \right>  }_{ \rm ren }\simeq H^{2}/24\pi^{2}$.
In the de Sitter spacetime, the renormalized vacuum fluctuation ${ \left< { \delta \phi   }^{ 2 } \right>  }_{ \rm ren }$ 
via the adiabatic regularization can be summarized as follows~\cite{Kohri:2016qqv}
\begin{equation}
\left< {  \delta \phi  }^{ 2 } \right>_{\rm ren}\simeq\begin{cases} { H }^{ 3 }t /4{ \pi  }^{ 2 } ,\quad\quad\quad  \left( m=0 \right) 
\\  3{ H }^{ 4} / 8{ \pi  }^{ 2 }m^{2},\quad\ \left( m\ll H \right) \\  H^{2}/24\pi^{2}. \ \quad\quad\  \left( m\gtrsim  H \right) \end{cases} \label{eq:klklksssg}
\end{equation}

Next, we discuss the renormalized vacuum fluctuation by using the point-splitting regularization,
which is the method of regularizing divergences as the point separation of $x$ and $x'$
in the two-point function~\cite{birrell1984quantum,bunch1978quantum,Vilenkin:1982wt}
and we simply show that the result of the point-splitting regularization is equivalent to the adiabatic regularization.
By using the point-splitting regularization, the renormalized vacuum fluctuation
${ \left< { \delta \phi   }^{ 2 } \right>  }_{ \rm ren }$  in the de Sitter spacetime
can be given as follows~\cite{Vilenkin:1982wt}
\begin{eqnarray}
\left< {  \delta \phi  }^{ 2 } \right>_{\rm ren}  = \frac { 1 }{ 16{ \pi  }^{ 2 } }\left[ { m }^{ 2 }+\left( \xi -\frac { 1 }{ 6 }  \right) R \right] \left[ \ln { \left( \frac { R }{  { 12m }^{ 2 } }  \right) +\psi \left( \frac { 3 }{ 2 } +\nu  \right)  } +\psi \left( \frac { 3 }{ 2 } -\nu  \right)  \right]
\label{eq:klkrkkdg}\nonumber,
\end{eqnarray}
where  $\nu=\sqrt{9/4-m^{2}H^{2}}$ and 
$\psi  \left( z \right)=\Gamma' \left( z \right)/\ \Gamma \left( z \right)$ is the digamma function. 
In the massless limit $m\rightarrow 0$, we can obtain the well-know expression as the following
\begin{equation}
\left< {  \delta \phi   }^{ 2 } \right>_{\rm ren}  \rightarrow\frac { R^{2} }{ 384{ \pi  }^{ 2 }m^{2} }=\frac { 3H^{4} }{ 8{ \pi  }^{ 2 }m^{2} }\label{eq:hl;lhf:kdg},
\end{equation}
which is equivalent to Eq.~(\ref{eq:klklksssg}). 
In the massive case $m\gtrsim H$
\footnote{
The digamma function $\psi  \left( z \right)$ for $z\gtrsim1$ can be approximated as follows~\cite{bunch1978quantum}
\begin{equation}
{\rm Re} \ {\psi  \left( \frac{3}{2}+i z \right)}= \log { z } +\frac { 11 }{ 24{ z }^{ 2 } } 
-\frac { 127 }{ 960{ z }^{ 4 } } +\cdots \label{eq:flklkdg}.\end{equation}
For the massive case $m\gtrsim H$, the logarithm terms can be approximately given as 
\begin{eqnarray}
\ln { \left( \frac { { H }^{ 2 } }{ { m }^{ 2 } }  \right)  } +\psi \left( \frac { 3 }{ 2 } +\nu  \right) +\psi \left( \frac { 3 }{ 2 } -\nu  \right)
\approx && \ln { \left( \frac { { H }^{ 2 } }{ { m }^{ 2 } }  \right)  } +\psi \left( \frac { 3 }{ 2 } +i\frac { m }{ H }  \right) +\psi \left( \frac { 3 }{ 2 } -i\frac { m }{ H }  \right),\\
\approx &&\frac { 11 }{ 12 } \frac { { H }^{ 2 } }{ { m }^{ 2 } } -\frac { 127 }{ 480 } \frac { { H }^{ 4 } }{ { m }^{ 4 } } +\cdots
\label{eq:dlkglkdg}.
\end{eqnarray}},
the renormalized vacuum fluctuation ${ \left< { \delta \phi   }^{ 2 } \right>  }_{ \rm ren }$ 
of the massive scalar field is written as follows:
\begin{eqnarray}
\left< {  \delta \phi  }^{ 2 } \right>_{\rm ren}  = \frac { 1 }{ 16{ \pi  }^{ 2 } }\left[ { m }^{ 2 }+\left( \xi -\frac { 1 }{ 6 }  \right) R \right] \left(\frac { 11 }{ 12 } \frac { { H }^{ 2 } }{ { m }^{ 2 } } -\frac { 127 }{ 480 } \frac { { H }^{ 4 } }{ { m }^{ 4 } } +\cdots  \right)
= \mathcal{O}\left(H^{2}\right)\label{eq:hldfkldg},
\end{eqnarray}
which corresponds with Eq.~(\ref{eq:klklksssg}).
Therefore, the renormalized vacuum fluctuations $\left< { \delta \phi    }^{ 2 } \right>_{\rm ren}$ 
via the point-splitting regularization are clearly equivalent to the results of Eq.~(\ref{eq:klklksssg}).
The curved background generates the classical vacuum field fluctuations, which contribute to
the backreaction effects of the classical and coherent scalar field $\phi$.
Therefore, we shift the scalar field $\phi^{2}\rightarrow \phi^{2}+\left< {  \delta \phi  }^{ 2 } \right>_{\rm ren} $
and then the effective scalar potential $V_{\rm eff}\left(\phi\right)$ can be written as follows
\begin{equation}
V_{\rm eff}\left(\phi\right)=
\frac{1}{2}\left(m^{2}+\xi R\right)\phi^{2}\rightarrow \frac{1}{2}\left(m^{2}+\xi R\right)\left(\phi^{2}+\left< {  \delta \phi  }^{ 2 } \right>_{\rm ren} \right).
\end{equation}
Therefore, the vacuum field fluctuation can generate the dynamical vacuum energy as
\begin{equation}
\rho_{\rm vacuum}=
\frac{1}{2}\left(m^{2}+\xi R\right)\left< {  \delta \phi  }^{ 2 } \right>_{\rm ren}.
\end{equation}
For the massive field case or the small Hubble scale $m\gtrsim H$, we can obtain 
the dynamical vacuum energy predicted by the running vacuum energy scenario as the follows:
\begin{eqnarray}
\rho_{\rm vacuum}\simeq
\frac{1}{2}\left(m^{2}+\xi R\right)\mathcal{O}\left(H^{2}\right)
\simeq m^{2}H^{2}. 
\end{eqnarray}
If there are super-Planckian massive fields as $m\gtrsim M_{\rm Pl}$, the dynamical vacuum energy 
exceeds the current dark energy as ${ \rho  }_{ \rm dark }\simeq 2.5\times10^{-47}\ {\rm GeV^{4}}$
and is inconsistent with the cosmological observation.
In the above discussion, we simply show that the curved background can generate 
the dynamical vacuum energy density as ${ \rho  }_{ \rm vacuum }\simeq m^{2}H^{2}$.
However, formally, we must consider the vacuum energy density in terms of the energy momentum tensor in curved spacetime. 
Following the literature~\cite{bunch1978quantum,birrell1978application,Bunch:1980vc,birrell1984quantum,Vilenkin:1982wt,Elias:2015yta},
let us consider the renormalized energy momentum tensor in curved spacetime
\footnote{
Note that the renormalization of the energy momentum tensor ${ T }_{ \mu \nu }$ in
  the curved background has some ambiguities depending on choices of
  the regularization and the vacuum
  states~\cite{Habib:1999cs,Anderson:2000wx,Anderson:2013ila,
  Anderson:2013zia,Markkanen:2016aes,Markkanen:2016jhg,Christensen:1977jc,Candelas:1980zt}. }.
The energy momentum tensor can be given as follows~\cite{Bunch:1980vc}
\begin{eqnarray}
\begin{split}
{ T }_{ \mu \nu  }&=\left( 1-2\xi  \right) { \partial  }_{ \mu  }\phi { \partial  }_{ \nu  }\phi +\left( 2\xi -\frac { 1 }{ 2 }  \right) { g }_{ \mu \nu  }{ { \partial  }^{ \mu  }\phi \partial  }_{ \mu  }\phi -2\xi \phi \nabla { \partial  }_{ \nu  }\phi \\ &+2\xi { g }_{ \mu \nu  }{ \phi \Box \phi -\xi { G }_{ \mu \nu  }{ \phi  }^{ 2 }+\frac { 1 }{ 2 } { m }^{ 2 }{ g }_{ \mu \nu  }{ \phi  }^{ 2 } },
\end{split}
\end{eqnarray}
where the diagonal and non-vanishing components are ${ T }_{ 00 }$ and 
${ T }_{11 }={ T }_{ 22 } ={ T }_{ 33 }$, which are the spatial components.
For the sake of convenience, we introduce the trace of the energy momentum tensor ${ T }^{\alpha }_{ \alpha }$ 
to satisfy the relation ${ T }_{ ii } = 1/3\left(  { T }_{ 00 }-a^{2}{ T }^{\alpha }_{ \alpha }\right)$.
Thus, the vacuum expectation values $\left< { T }_{ \mu \nu  } \right>$ of the energy momentum tensor 
for the mode function ${ \delta\chi  }_{ k }\left(\eta\right)$ can be given by
\begin{eqnarray}
\begin{split}
\left< { T }_{ 00 } \right> &=\frac { 1 }{ 4{ \pi  }^{ 2 }C\left( \eta  \right)  } \int { dk{ k }^{ 2 } } \biggl[ { \left| \delta { \chi ' }_{ k } \right|  }^{ 2 }
+{ \omega  }_{ k }^{ 2 }{ \left| \delta { \chi  }_{ k } \right|  }^{ 2 } \\ &  +\left( \xi -\frac { 1 }{ 6 }  \right) 
\left( 3D\left( \delta \chi _{ k }\delta { \chi  }_{ k }^{ *' }+\delta { \chi  }_{ k }^{ * }{ \delta \chi ' }_{ k } \right) -\frac { 3 }{ 2 } { D }^{ 2 }{ \delta { \chi  } }_{ k }^{ 2 } \right)  \biggr],
\end{split}
\end{eqnarray}
\begin{eqnarray}
\begin{split}
\left< { T }^{\alpha }_{ \alpha } \right> =\frac { 1 }{ 2{ \pi  }^{ 2 }{ C }^{ 2 }\left( \eta  \right)  }& \int { dk{ k }^{ 2 } } 
 \biggl[Cm^{2}{ \left| \delta { \chi  }_{ k } \right|  }^{ 2 }+6\left( \xi -\frac { 1 }{ 6 }\right)  \biggl( 
 { \left| \delta { \chi  }_{ k } \right|  }^{ 2 }-\frac{1}{2}D\left( \delta \chi _{ k }\delta { \chi  }_{ k }^{ *' }
+\delta { \chi  }_{ k }^{ * }{ \delta \chi ' }_{ k }\right) \\ & -{ \omega  }_{ k }^{ 2 }{ \left| \delta { \chi  }_{ k } \right|  }^{ 2 }
-\frac{1}{2}D'{ \left| \delta { \chi  }_{ k } \right|  }^{ 2 }-\left( \xi -\frac { 1 }{ 6 }  \right)\left(3D'+\frac{3}{2}D^{2}\right){ \left| \delta { \chi  }_{ k } \right|  }^{ 2 }
 \biggr)  \biggr],
\end{split}
\end{eqnarray}
where the energy momentum tensor express the energy density as $\rho=\left< { T }_{ 00 } \right>/C$ and the 
pressure as $p=\left< { T }_{ ii } \right>/C$.
As previously discussed in the vacuum field fluctuations, we use the adiabatic (WKB) approximation
to the mode ${ \delta\chi  }_{ k }\left(\eta\right)$, and therefore, 
the vacuum expectation values $\left< { T }_{ \mu \nu  } \right>$ of the energy momentum tensor
can be given by the adiabatic approximation~\cite{Bunch:1980vc}
\begin{eqnarray}
\begin{split}
\left< { T }_{ 00 } \right> =&\frac { 1 }{ 8{ \pi  }^{ 2 }C\left( \eta  \right)  } \int { dk{ k }^{ 2 } } \biggl[ 
2{ \omega  }_{ k }+\frac{C^{2}m^{4}D^{2}}{16{ \omega  }_{ k }^{ 5}}
-\frac{C^{2}m^{4}}{64{ \omega  }_{ k }^{ 7}}\left(2D''D-D'^{2}+4D'D^{2}+D^{4}\right) \\ &  
+\frac{7C^{3}m^{6}}{64{ \omega  }_{ k }^{ 9}}\left(D'D^{2}+D^{4}\right)-\frac{105C^{4}m^{8}D^{4}}{1024{ \omega  }_{ k }^{11}} \\ &  
+\left( \xi -\frac { 1 }{ 6 }  \right)  \biggl( -\frac{3D^{2}}{2{ \omega  }_{ k }}-\frac{3Cm^{2}D^{2}}{2{ \omega  }_{ k }^{3}}+
\frac{Cm^{2}}{8{ \omega  }_{ k }^{5}}\left(6D''D-3D'^{2}+6D'D^{2}\right) \\ & 
-\frac{C^{2}m^{4}}{64{ \omega  }_{ k }^{7}}\left(120D'D^{2}+105D^{4}\right)+\frac{105C^{3}m^{6}D^{4}}{64{ \omega  }_{ k }^{9}} \biggr)  \\ &  
+\left( \xi -\frac { 1 }{ 6 }  \right)^{2} \left( -\frac{1}{16{ \omega  }_{ k }^{3}}\left(72D''D-36D'^{2}-27D^{4}\right) 
+\frac{Cm^{2}}{8{ \omega  }_{ k }^{5}}\left(54D'D^{2}+27D^{4}\right)\right)  \biggr],
\end{split}
\end{eqnarray}
\begin{eqnarray}
\begin{split}
\left< { T }^{\alpha }_{ \alpha } \right> =&\frac { 1 }{ 8{ \pi  }^{ 2 }C^{2}\left( \eta  \right)  } \int { dk{ k }^{ 2 } } \biggl[ 
\frac{Cm^{2}}{{ \omega  }_{ k }}+\frac{C^{2}m^{4}}{8{ \omega  }_{ k }^{5}}\left(D'+D^{2}\right)-\frac{5C^{3}m^{6}D^{2}}{32{ \omega  }_{ k }^{7}} \\ &
-\frac{C^{2}m^{4}}{32{ \omega  }_{ k }^{7}}\left(D'''+4D''D+3D'^{2}+6D'D^{2}+D^{4}\right)  \\ &
+\frac{C^{3}m^{6}}{128{ \omega  }_{ k }^{9}}\left(28D''D+21D'^{2}+126D'D^{2}+49D^{4}\right)
 -\frac{231C^{4}m^{8}}{256{ \omega  }_{ k }^{11}}\left(D'D^{2}+D^{4}\right)  \\ &
+\frac{1155C^{5}m^{10}D^{4}}{2048{ \omega  }_{ k }^{13}}+ \left( \xi -\frac { 1 }{ 6 }  \right)
\biggl( -\frac{3D'}{{ \omega  }_{ k }}-\frac{Cm^{2}}{{ \omega  }_{ k }^{3}}\left(3D'+\frac{3}{4}D^{2}\right) \\ &
+\frac{9C^{2}m^{4}D^{2}}{4{ \omega  }_{ k }^{5}}+\frac{Cm^{2}}{4{ \omega  }_{ k }^{5}}\left(3D'''+6D''D+\frac{9}{2}D'^{2}+3D'D^{2}\right) \\ &
-\frac{C^{2}m^{4}}{32{ \omega  }_{ k }^{7}}\left(120D''D+90D'^{2}+390D'D^{2}+105D^{4}\right) \\ &
+\frac{C^{3}m^{6}}{128{ \omega  }_{ k }^{9}}\left(1680D'D^{2}+1365D^{4}\right)- \frac{945C^{4}m^{8}D^{4}}{128{ \omega  }_{ k }^{11}} \biggr)  
+\left( \xi -\frac { 1 }{ 6 }  \right)^{2} \biggl( - \frac{1}{4{ \omega  }_{ k }^{3}}\left(18D'''-27D'D^{2}\right)  \\ &
+\frac{C m^{2}}{32{ \omega  }_{ k }^{5}}\left(432D''D+324D'^{2}+648D'D^{2}+27D^{4}\right)-\frac{C^{2}m^{4}}{16{ \omega  }_{ k }^{7}}
\left(270D'D^{2}+135D^{4} \right)\biggr) \biggr].
\end{split}
\end{eqnarray}
To renormalize the energy momentum tensor of quantum fields in curved spacetime, 
we must consider at least the forth-order adiabatic approximation 
(see, e.g. Ref.\cite{Bunch:1980vc} for the details).
However, as already mentioned in the vacuum field fluctuations 
${ \left< { \delta \phi   }^{ 2 } \right>  }$, the high-order adiabatic terms are finite and 
the divergence terms of the energy momentum tensor originate from the low-order adiabatic approximation
\begin{align}
\left< { T }_{ 00 } \right>_{\rm low-order}=\ &\frac { 1 }{ 8{ \pi  }^{ 2 }C\left( \eta  \right)  } \int { dk{ k }^{ 2 } } \biggl[ 
2{ \omega  }_{ k }-\frac{3}{2}D^{2}
\left( \xi -\frac { 1 }{ 6 }  \right)  \left( \frac{1}{{ \omega  }_{ k }}+\frac{Cm^{2}}{{ \omega  }_{ k }^{3}}\right)\nonumber  \\ &  
-\frac{1}{16{ \omega  }_{ k }^{3}}\left( \xi -\frac { 1 }{ 6 }  \right)^{2} \left(72D''D-36D'^{2}-27D^{4}\right)  \biggr],
\end{align}
\begin{align}
\left< { T }^{\alpha }_{ \alpha } \right>_{\rm low-order} =\ &\frac { 1 }{ 8{ \pi  }^{ 2 }C^{2}\left( \eta  \right)  } \int { dk{ k }^{ 2 } } \biggl[ 
\frac{Cm^{2}}{{ \omega  }_{ k }} - \left( \xi -\frac { 1 }{ 6 }  \right)
\left( \frac{3D'}{{ \omega  }_{ k }}
+\frac{Cm^{2}}{{ \omega  }_{ k }^{3}}\left(3D'+\frac{3}{4}D^{2}\right)   \right) \nonumber  \\ &
- \frac{1}{4{ \omega  }_{ k }^{3}}\left( \xi -\frac { 1 }{ 6 }  \right)^{2}\left(18D'''-27D'D^{2}\right)  \biggr].
\end{align}
By using the dimensional regularization
\footnote{
The divergent momentum integrals can be simplified as
\begin{equation}
I\left( 0,n \right) =\int { \frac { { d }^{ 3 }k }{ { \left( 2\pi  \right)  }^{ 3 } } \frac { 1 }{ { \omega  }_{ k }^{ n } } 
=\int { \frac { { d }^{ 3 }k }{ { \left( 2\pi  \right)  }^{ 3 } } \frac { 1 }{ \left( { k }^{ 2 }+{ a }^{ 2 }{ m }^{ 2 } \right) ^{ n/2 } } }  }.
\end{equation}
We can regulate these integrals of the spatial dimensions $ 3-2\epsilon $ as 
\begin{equation}
I\left( \epsilon,n \right) =\int { \frac { { d }^{ 3-2\epsilon }k }{ { \left( 2\pi  \right)  }^{3-2\epsilon } } }
\frac { \left(a\mu\right)^{2\epsilon} }{ { \omega  }_{ k }^{ n } }=\frac { { \left( am \right)  }^{ 3-n }}
{ 8{ \pi  }^{ 3/2 } }
\frac { \Gamma \left( \epsilon -\frac { 3-n }{ 2 }  \right)  }{ \Gamma \left( \frac { n }{ 2 }  \right)  }
{ \left( \frac { 4\pi { \mu  }^{ 2 } }{ { m }^{ 2 } }  \right)  }^{ \epsilon  }.
\end{equation}},
The low-order and unphysical divergence terms of the energy momentum tensor
can be written by
\begin{eqnarray}
\begin{split}
\left< { T }_{ 00 } \right>_{\rm low-order} =&-\frac { { m }^{ 4 }C}{ 64{ \pi  }^{ 2 } } \left[ \frac { 1 }{ \epsilon  } 
+\frac { 3 }{ 2 } -\gamma +\ln { 4\pi  } +\ln { \frac { { \mu  }^{ 2 } }{ { m }^{ 2 } }  }  \right]  \\ &
- \frac { 3{ m }^{ 2 }{ D }^{ 2 } }{ 32{ \pi  }^{ 2 }} \left( \xi -\frac { 1 }{ 6 }  \right) 
\left[ \frac { 1 }{ \epsilon  } +\frac { 1 }{ 2 } -\gamma +\ln { 4\pi  } +\ln { \frac { { \mu  }^{ 2 } }{ { m }^{ 2 } }  }  \right]  \\ &
-\frac { 1 }{ 256{ \pi  }^{ 2 }{ C } } { \left( \xi -\frac { 1 }{ 6 }  \right)  }^{ 2 }\left(72D''D-36D'^{2}-27D^{4}\right)
\left[ \frac { 1 }{ \epsilon  }-\gamma +\ln { 4\pi  } +\ln { \frac { { \mu  }^{ 2 } }{ { m }^{ 2 } }  }  \right].
\end{split}
\end{eqnarray}
The high-order adiabatic terms of the energy momentum tensor are 
finite and provide the dynamical contributions on the curved background
\begin{eqnarray}
\begin{split}
\left< { T }_{ 00 } \right>_{\rm high-order} =\ &\frac { { m }^{ 2 }{ D }^{ 2 } }{ 384{ \pi  }^{ 2 } }-\frac { 1 }{ 2880{ \pi  }^{ 2 }{ C }} 
\left( \frac { 3 }{ 2 } D''D-\frac { 3 }{ 4 } { D' }^{ 2 }-\frac { 3 }{ 8 } { D }^{ 4 } \right)  \\ &
+\frac { 1 }{ 256{ \pi  }^{ 2 }{ C } } \left( \xi -\frac { 1 }{ 6 }  \right) \left( 8D''D-4{ D' }^{ 2 }-3{ D }^{ 4 } \right) \\ &
+ \frac { 1 }{ 64{ \pi  }^{ 2 }{ C } } \left( \xi -\frac { 1 }{ 6 }  \right) ^{ 2 }\left( 18D'D^{ 2 }+9{ D }^{ 4 } \right) .
\end{split}
\end{eqnarray}
Thus, the vacuum expectation values of the energy momentum tensor in curved spacetime
can be given by the divergent low-order terms and the finite high-order terms,
\begin{eqnarray}
{ \left< { T }_{ \mu \nu  } \right>  }=
{ \left< { T }_{ \mu \nu  } \right>  }_{ \rm low-order }+{ \left< { T }_{ \mu \nu  } \right>  }_{ \rm high-order }.
\end{eqnarray}
The unphysical divergence of the energy momentum tensor can be removed by the subtraction of the 
bare gravitational coupling constants of the Einstein equations~\cite{birrell1984quantum}
\begin{equation}
\frac { 1 }{ 8\pi { G }_{ N } } { G }_{ \mu \nu  }+{ \rho  }_{ \Lambda  } { g }_{ \mu \nu  }+{ a }_{ 1 }{ H }_{ \mu \nu  }^{ \left( 1 \right)  }
{ +a }_{ 2 }{ H }_{ \mu \nu  }^{ \left( 2 \right)  }{ +a }_{ 3 }{ H }_{ \mu \nu  }=\left< { T }_{ \mu \nu  } \right> .
\end{equation}
Now, we divide the low-order adiabatic terms into the divergent terms and the finite terms as the following
\begin{equation}
\left< { T }_{ \mu \nu  } \right>=  \left< { T }_{ \mu \nu  } \right>_{\rm div}+\left< { T }_{ \mu \nu  } \right>_{\rm finite}
+\left< { T }_{ \mu \nu  } \right>_{\rm high-order}.
\end{equation}
In order to renormalize the divergences of the energy momentum tensor, we divide the 
bare gravitational couplings ${ G }_{ N }$, ${ \rho  }_{ \Lambda  }$, ${ a }_{ 1}$ 
into the renormalized terms and the counter terms as
\begin{eqnarray}
{ G }_{ N }&=&{ G }_{ N }\left( \mu  \right) +\delta{ G }_{ N }, \\
{ \rho  }_{ \Lambda  } &=&{ \rho  }_{ \Lambda  }\left( \mu  \right)+\delta{ \rho  }_{ \Lambda  }, \\
{ a }_{ 1}&=&{ a }_{ 1 } \left( \mu  \right)+\delta { a }_{ 1 }.
\end{eqnarray}
The  divergences of the energy momentum tensor can be absorbed by the counter terms of the gravitational couplings 
$\delta{ G }_{ N }$, $\delta{ \rho  }_{ \Lambda  }$, $\delta{ a }_{ 1}$ as follows:
\begin{eqnarray}
\begin{split}
 \left< { T }_{ 00  } \right>_{\rm div}&=
\frac { 1 }{ 8\pi \delta{ G }_{ N } } { G }_{ 00  }+\delta{ \rho  }_{ \Lambda  }{ g }_{ 00  }+\delta{ a }_{ 1 }{ H }_{ 00  }^{ \left( 1 \right)  }
+\delta{ a }_{ 2 }{ H }_{ 00  }^{ \left( 2 \right)  }+\delta{ a }_{ 3 }{ H }_{ 00 } \\ &
=\frac { 1 }{ 8\pi \delta{ G }_{ N } }\left( -\frac { 3 }{ 4 }{ D }^{ 2 }   \right) + \delta{ \rho  }_{ \Lambda  } \left( C \right)
+ \delta{ a }_{ 1 } \left( \frac{-72D''D+36D'^{2}+27D^{4}}{8C} \right)+\cdots.
\end{split}
\end{eqnarray}
Therefore, the following conditions of the gravitational counter terms
$\delta{ G }_{ N }$, $\delta{ \rho  }_{ \Lambda  }$, $\delta{ a }_{ 1}$ can be imposed as
\begin{eqnarray}
\delta{ \rho  }_{ \Lambda  }&=& \frac { { m }^{ 4 } }{ 64{ \pi  }^{ 2 } } \left[ \frac { 1 }{ \epsilon  } -\gamma +\ln { 4\pi  }  \right],   \\
\frac { 1 }{ 8\pi \delta{ G }_{ N } } &=& -\frac { m^{2} }{ 8\pi^{2} }\left( \xi -\frac { 1 }{ 6 }  \right)  
\left[ \frac { 1 }{ \epsilon  } -\gamma +\ln { 4\pi  }   \right], \\
\delta a_{1}&=&-\frac{1}{32\pi^{2}}\left( \xi -\frac { 1 }{ 6 }  \right)^{2}
\left[ \frac { 1 }{ \epsilon  } -\gamma +\ln { 4\pi  } \right].
\end{eqnarray}
Thus, we can obtain the renormalized Einstein's equations as follows
\begin{equation}
\frac { 1 }{ 8\pi { G }_{ N }  \left( \mu  \right) } { G }_{ \mu \nu  }+{ \rho  }_{ \Lambda  }\left( \mu  \right) 
 { g }_{ \mu \nu  }+{ a }_{ 1 } \left( \mu  \right) { H }_{ \mu \nu  }^{ \left( 1 \right)  }
+{ a }_{ 2 } \left( \mu  \right) { H }_{ \mu \nu  }^{ \left( 2 \right)  }+{a }_{ 3 } \left( \mu  \right) { H }_{ \mu \nu  }
=\left< { T }_{ \mu \nu  } \right>_{\rm ren},
\end{equation}
where the renormalized energy momentum tensor is given by
\begin{eqnarray}
\left< { T }_{\mu \nu } \right>_{\rm ren} =\left< { T }_{ \mu \nu } \right>_{\rm finite}+\left< { T }_{\mu \nu } \right>_{\rm high-order}.
\end{eqnarray}
The renormalized finite energy momentum tensor of $\left< { T }_{ 00 } \right>_{\rm finite}$ can be given by
\begin{eqnarray}
\begin{split}
\left< { T }_{ 00 } \right>_{\rm finite} =\ &\frac { { m }^{ 4 }C}{ 64{ \pi  }^{ 2 } }
\left(\ln { \frac { { m  }^{ 2 } }{ { \mu }^{ 2 } }  }-\frac{3}{2}\right)  +\frac { 3{ m }^{ 2 }{ D }^{ 2 } }{ 32{ \pi  }^{ 2 }} \left( \xi -\frac { 1 }{ 6 }  \right) 
\left(\ln { \frac { { m  }^{ 2 } }{ { \mu }^{ 2 } }  }-\frac{1}{2}\right) \\ &
+\frac { 1 }{ 256{ \pi  }^{ 2 }{ C } } { \left( \xi -\frac { 1 }{ 6 }  \right)  }^{ 2 }\left(72D''D-36D'^{2}-27D^{4}\right)
\left(\ln { \frac { { m  }^{ 2 } }{ { \mu }^{ 2 } }  }\right).
\end{split}
\end{eqnarray}
Thus, the renormalized vacuum energy density can be given by the renormalized 
energy momentum tensor as follows:
\begin{eqnarray}
\begin{split}
\rho_{\rm vacuum }=&\left< { T }_{ 00 } \right>_{\rm ren}/C \\
=\ &\frac { { m }^{ 4 }}{ 64{ \pi  }^{ 2 } }
\left(\ln { \frac { { m  }^{ 2 } }{ { \mu }^{ 2 } }  }-\frac{3}{2}\right)  +\frac { 3{ m }^{ 2 }{ D }^{ 2 } }{ 32{ \pi  }^{ 2 }C} \left( \xi -\frac { 1 }{ 6 }  \right) 
\left(\ln { \frac { { m  }^{ 2 } }{ { \mu }^{ 2 } }  }-\frac{1}{2}\right) \\ &
+\frac { 1 }{ 256{ \pi  }^{ 2 }{ C^{2} } } { \left( \xi -\frac { 1 }{ 6 }  \right)  }^{ 2 }\left(72D''D-36D'^{2}-27D^{4}\right)
\left(\ln { \frac { { m  }^{ 2 } }{ { \mu }^{ 2 } }  }\right) \\&
+\frac { { m }^{ 2 }{ D }^{ 2 } }{ 384{ \pi  }^{ 2 }C }-\frac { 1 }{ 2880{ \pi  }^{ 2 }{ C^{2} }} 
\left( \frac { 3 }{ 2 } D''D-\frac { 3 }{ 4 } { D' }^{ 2 }-\frac { 3 }{ 8 } { D }^{ 4 } \right)  \\ &
+\frac { 1 }{ 256{ \pi  }^{ 2 }{ C^{2} } } \left( \xi -\frac { 1 }{ 6 }  \right) \left( 8D''D-4{ D' }^{ 2 }-3{ D }^{ 4 } \right) \\ &
+ \frac { 1 }{ 64{ \pi  }^{ 2 }{ C^{2} } } \left( \xi -\frac { 1 }{ 6 }  \right) ^{ 2 }\left( 18D'D^{ 2 }+9{ D }^{ 4 } \right) \label{eq:tutdfgdfsdg},
\end{split}
\end{eqnarray}
where the first parts originating from $\left< { T }_{ 00 } \right>_{\rm finite}$ 
depend on the renormalization scale, but,
the renormalization scale dependence can be canceled by the running gravitational couplings
$G_{N}$, ${ \rho  }_{ \Lambda  }$, ${a }_{ 1}$
in the renormalized Einstein's equations. On the other hand, the 
second parts originating from $\left< { T }_{ 00 } \right>_{\rm high-order}$ 
are independent of the renormalization scale and express 
the dynamical contributions of the curved background as $\rho_{\rm vacuum }\simeq m^{2}H^{2}/96\pi^{2}$
\footnote{
By using the point-splitting regularization, the renormalized vacuum energy density
in the de Sitter spacetime can be given by~\cite{bunch1978quantum,Christensen:1976vb,Christensen:1978yd}
\begin{eqnarray}
\begin{split}
{ \rho  }_{ \rm vacuum }&=\frac { 1 }{ 64{ \pi  }^{ 2 } } \Biggl\{ { m }^{ 2 }\left[ { m }^{ 2 }+\left( \xi -\frac{1}{6} \right) 12{ H }^{ 2 } \right]
 \left[ \ln { \left( \frac { { H}^{ 2 } }{ { m }^{ 2 } }  \right)  } + \psi \left( \frac { 3 }{ 2 } +\nu  \right) +\psi \left( \frac { 3 }{ 2 } -\nu  \right) \right] \\
&-{ m }^{ 2 }\left( \xi -\frac{1}{6} \right) 12{ H }^{ 2 }-\frac { 2 }{ 3 } { m }^{ 2 }{ H }^{ 2 }-\frac { 1 }{ 2 }\left( \xi -\frac{1}{6} \right) 12{ H }^{ 2 }
+\frac { { H }^{ 4 } }{ 15 }  \Biggr\} ,
\end{split} \nonumber
\end{eqnarray}
}.
In the matter dominated Universe as $a\left(\eta\right)=\eta^{2}/9$,
the renormalized vacuum energy density can be given by 
\begin{eqnarray}
\begin{split}
\rho_{\rm vacuum }
=&\frac { { m }^{ 4 }}{ 64{ \pi  }^{ 2 } }
\left(\ln { \frac { { m  }^{ 2 } }{ { \mu }^{ 2 } }  }-\frac{3}{2}\right)  +\frac { 3{ m }^{ 2 }}{ 8{ \pi  }^{ 2 }} \left( \xi -\frac { 1 }{ 6 }  \right) H^{2}
\left(\ln { \frac { { m  }^{ 2 } }{ { \mu }^{ 2 } }  }-\frac{1}{2}\right)  \\&
-\frac { 81 }{ 64{ \pi  }^{ 2 }} { \left( \xi -\frac { 1 }{ 6 }  \right)  }^{ 2 }H^{4}
\left(\ln { \frac { { m  }^{ 2 } }{ { \mu }^{ 2 } }  }\right)
+\frac { { m }^{ 2 }{ H }^{ 2 } }{ 96{ \pi  }^{ 2 } }
+\frac { H^{4} }{ 768{ \pi  }^{ 2 }} \\ & -\frac { 9 }{ 64{ \pi  }^{ 2 } } \left( \xi -\frac { 1 }{ 6 }  \right)H^{4} 
+ \frac { 9 }{ 8{ \pi  }^{ 2 }} \left( \xi -\frac { 1 }{ 6 }  \right) ^{ 2 }H^{4},
\end{split}
\end{eqnarray}
On the other hand, 
in the de Sitter Universe as $a\left(\eta\right)=-1/H\eta$, the renormalized vacuum energy density can be obtained by 
\begin{eqnarray}
\begin{split}
\rho_{\rm vacuum }
=&\frac { { m }^{ 4 }}{ 64{ \pi  }^{ 2 } }
\left(\ln { \frac { { m  }^{ 2 } }{ { \mu }^{ 2 } }  }-\frac{3}{2}\right)  +\frac { 3{ m }^{ 2 } }{ 8{ \pi  }^{ 2 }} \left( \xi -\frac { 1 }{ 6 }  \right) H^{2}
\left(\ln { \frac { { m  }^{ 2 } }{ { \mu }^{ 2 } }  }-\frac{1}{2}\right)  \\&
+\frac { { m }^{ 2 }{ H }^{ 2 } }{ 96{ \pi  }^{ 2 } }
-\frac { H^{4} }{ 960{ \pi  }^{ 2 }}
+ \frac { 9 }{ 2{ \pi  }^{ 2 }} \left( \xi -\frac { 1 }{ 6 }  \right) ^{ 2 }H^{4},
\end{split}
\end{eqnarray}
Thus, the dynamical vacuum energy has cosmological validity in the framework of QFT in curved spacetime,
and the outstanding cosmological influence on the observed Universe.
For instance, this fact leads to the trans-Planckian regulation that there would not exist 
trans-Planckian massive fields as $m \gtrsim  M_{\rm Pl}$
\footnote{
In the context of the Einstein Gravity, there is a naive conjecture 
to restrict the trans-Planckian physics.
If the Compton length of the particles is smaller than the Schwarzschild radius of the black hole (BH)
as $1/m <  2m^{2}/M_{\rm Pl}^{2}$,
they lose the particle pictures and we recognize the Planck mass $M_{\rm Pl}$
as the upper bound on the mass of any elementary particle.
However, for the Planckian mass regime $m\approx M_{\rm Pl}$, we should not recognize 
these particles as the classical black holes
due to the quantum fluctuations, and therefore, we can treat such particles as 
quantum black holes of the Planck-scale mass $\mathcal{O}\left(M_{\rm Pl}\right)$
and consider the quantum corrections of such quantum black holes~\cite{Dvali:2010bf,Dvali:2010ue}.
}.

\section{Physical interpretation of the RG running of the cosmological constant and the effective scalar potential}
\label{sec:significance}
In this section, we explore more specifically physical interpretation of the RG running effects
of the cosmological constant and the effective scalar potential.
These matters have been discussed considerably in the literature
\cite{Foot:2007wn,Shapiro:2008yu,Shapiro:2009dh,Sola:2011qr,Sola:2013gha,
Ward:2009wq,Ward:2014sla,Hamber:2013rb}, 
but here we discuss and review once again.
In particular, by considering the Coulomb effective potential, 
we clarify the theoretical definition of the renormalization scale under the effective scalar potential. 
In consequence, we can clearly obtain 
the physical interpretation of the RG running effects of the cosmological constant, and furthermore,
understand that the dynamical vacuum energy corresponds to the vacuum field fluctuations on
the curved background rather than the RG running of the cosmological constant.

In principle, the vacuum energy density as the ground state of the Universe can be determined by the minimum of the effective potential 
$V_{\rm eff}\left(\phi\right)$. Therefore, we revisit the renormalization-scale dependence of the vacuum energy  
in terms of the effective potential in flat spacetime and curved spacetime.
In the framework of QFT, the effective potential is derived from the effective action.
Here, the effective action in flat spacetime is defined via the functional Legendre transform
\begin{equation}
\Gamma \left[ \phi  \right] =W\left[ J \right] -\int { { d }^{ 4 }x J\left( x \right) \phi \left( x \right)  } ,
\end{equation}
where the source $J \left( x \right)$ is written as $J\left( x \right)= -\delta \Gamma \left[ \phi  \right]  / \delta \phi \left( x \right)$ and
the generating function $W\left[ J \right]$ is defined by the Feynman path integral
\begin{equation}
{ e }^{ iW\left[ J \right]  }
\equiv \int { \mathcal{D}\phi  }\ {\rm exp}\left\{ i\int { { d }^{ 4 }x \left( \mathcal{L}_{\phi} +J\phi  \right)  }  \right\} ,
\end{equation}
where $ \mathcal{L}_{\phi}$ is the classic Lagrangian for the scalar field and 
$\phi$ is the classical scalar field in flat spacetime. For simplicity, we consider the following Lagrangian
\begin{equation}
\mathcal{L}_{\phi}=\frac { 1 }{ 2 } { \partial  }_{ \mu  }\phi { \partial  }^{ \mu  }\phi -\frac { 1 }{ 2 } { m }^{ 2 }{ \phi  }^{ 2 }-\frac { 1 }{ 4 } { \lambda  }{ \phi  }^{ 4 }-{ \rho  }_{ \Lambda  }.
\end{equation}
The general form of the effective action consists of the standard kinetic term multiplied 
by the wave-function renormalization factor $Z_{\rm eff}\left[ \phi  \right]$,
the effective potential $V_{\rm eff}\left(\phi\right)$ and the higher derivative terms as follows
\begin{equation}
\Gamma \left[ \phi  \right] =\int { { d }^{ 4 }x\left( Z_{\rm eff}\left[ \phi  \right]
{ \partial  }_{ \mu  }\phi { \partial  }^{ \mu  }\phi- V_{\rm eff}\left(\phi\right)+\cdots \right)  }.
\end{equation}
The one-loop effective potential in flat spacetime is given by
\begin{eqnarray}
V_{\rm eff}\left( \phi  \right) =\frac{1}{2}m^{2}\phi^{2}+\frac{\lambda}{4}\phi^{4}+{ \rho  }_{ \Lambda  }
+\frac { { \left( { m }^{ 2 }+3\lambda \phi^{2} \right) }^{ 2 } }{ 64{ \pi  }^{ 2 } }
\left( \ln {\frac { { m }^{ 2 }+3\lambda \phi^{2} }{ { \mu  }^{ 2 } }} - \frac{3}{2} \right),\label{eq:tutusssdg}
\end{eqnarray}
where $\mu$ is the renormalization scale and 
${ \rho  }_{ \Lambda  }$ is the cosmological constant which depends on the scale $\mu$.
However, it is clear that the $\mu$-dependence of the running coupling is exactly canceled 
via the log-corrections and the effective potential is formally renormalization scale invariant as follows:
\begin{eqnarray}
\frac { d }{ d\ln{\mu}  }{ V }_{\rm eff }\left( \phi ,{ m }^{ 2 },\lambda ,{ \rho  }_{ \Lambda  }, \mu  \right) =0 .
\end{eqnarray}
Thus, the effective potential satisfies the renormalized group equation (RGE) as 
\begin{equation}
\left( \mu \frac { \partial  }{ \partial \mu  } +{ \beta  }_{ \lambda  }\frac { \partial  }{ \partial \lambda   } 
+{ \beta  }_{ m^{2}  }\frac { \partial  }{ \partial { m }^{ 2 } } -{ \gamma  }_{ \phi  }\phi \frac { \partial  }{ \partial \phi  } +{ \beta  }_{ { \rho  }_{ \Lambda  } }\frac { \partial  }{ \partial { \rho  }_{ \Lambda  } }  \right)
{ V }_{ \rm eff }\left( \phi ,{ m }^{ 2 },\lambda ,{ \rho  }_{ \Lambda  }, \mu  \right)=0 .
\end{equation}
where ${ \beta  }_{ { \lambda  }, m^{2}, { \rho  }_{ \Lambda  } }\left(\mu\right)$ are the $\beta$ functions of the coupling 
${ \lambda  }\left(\mu\right)$, $m^{2}\left(\mu\right)$, ${ \rho  }_{ \Lambda  } \left(\mu\right)$ and
$\gamma_{\phi}$ is the $\gamma$ function of the scalar field $\phi$.
Formally, the effective potential is written as the renormalization scale independent form.
Therefore, we may consider any value of the renormalization scale $\mu$.
Thus, the minimum of the effective potential is also renormalization scale independent and 
consequently the vacuum energy of the effective potential 
does not run with any renormalization scale in the Minkowski spacetime.
This situation is the same as the renormalized vacuum energy and
the scale dependence of the cosmological constant is hidden from the definition of the effective potential.

Next, let us consider the effective potential in curved spacetime \cite{Elizalde:1993ee,Elizalde:1993ew,
Elizalde:1993qh,Elizalde:1994gv,Elizalde:1995at}.
The QFT in curved spacetime, which is the standard extension of QFT from Minkowski spacetime to curved spacetime,
describes the quantum effects of the time-dependent gravitational background
(see Ref.\cite{parker2009quantum,birrell1982p,fulling1989aspects,DeWitt:1975ys,Buchbinder:1992rb,Bunch:1979uk,Shapiro:2008sf}
for the general reviews).
For simplicity, we consider the following Lagrangian with the non-minimal coupling $\xi$ as 
\begin{equation}
\mathcal{L}_{\phi}=\frac { 1 }{ 2 }g^{\mu\nu}{ \partial  }_{ \mu  }\phi { \partial  }_{ \nu  }\phi -\frac { 1 }{ 2 } 
\left({ m }^{ 2 }+\xi R \right){ \phi  }^{ 2 }-\frac { 1 }{ 4 } { \lambda  }{ \phi  }^{ 4 }-{ \rho  }_{ \Lambda  }.
\end{equation}
However, in order to have a renormalizable theory in curved spacetime, 
we must consider the following gravitational Lagrangian including the high-order derivative as
\begin{equation}
\mathcal{L}_{R}={ \rho  }_{ \Lambda  } +\kappa R+{ a }_{ 1 }{ R }^{ 2 }+{ a }_{ 2 }{ C }^{ 2 }
+{ a }_{ 3 }E+{ a }_{ 4 }\Box R ,
\end{equation}
where $\kappa$ is the inverse of the Newton's gravitational constant $\kappa=1/16\pi G_{N}$,
$a_{1},a_{2},a_{3},a_{4}$ are the gravitational coupling constants, ${ C }^{ 2 }$ is the square of the Weyl tensor and
$E$ is the Gauss-Bonnet term.
Thus, the total Lagrangian is given by
\begin{equation}
\mathcal{L}_{\rm total}=\mathcal{L}_{\phi}+\mathcal{L}_{R}+{ a }_{ 5 }\Box \phi^{2} ,
\end{equation}
where we add the term ${ a }_{ 5 }\Box \phi^{2}$ to the total Lagrangian $\mathcal{L}_{\rm total}$
(see Ref.\cite{Buchbinder:1992rb} for the details) and it is well known that the above Lagrangian is renormalizable in curved spacetime. 
Therefore, we can obtain the effective action  and the effective potential in curved spacetime via
the total Lagrangian $\mathcal{L}_{\rm total}$.
In the same way as the flat spacetime, the effective potential in curved spacetime satisfies the standard RGE as follows
\begin{equation}
\left( \mu \frac { \partial  }{ \partial \mu  } +{ \beta  }_{ { \lambda  }_{ i } }\frac { \partial  }{ \partial { \lambda  }_{ i } }
 -\gamma_{\phi} \phi \frac { \partial  }{ \partial \phi  }  \right)V_{\rm eff}\left(\phi, { \lambda  }_{ i },\mu \right) =0 ,
\end{equation}
where ${ \lambda  }_{ i }\left(\mu\right)=\lambda\left(\mu\right), \xi\left(\mu\right),
m^{2}\left(\mu\right), { \rho  }_{ \Lambda  }\left(\mu\right)$ 
express coupling constants,
${ \beta  }_{ { \lambda  }_{ i } }\left(\mu\right)$ are the $\beta$ functions of the coupling 
${ \lambda  }_{ i }\left(\mu\right)$ and $\gamma_{\phi}$ is the $\gamma$ function of the scalar field $\phi$.
The one-loop $\beta$ functions in curved spacetime are given as follows~\cite{Elizalde:1993ee,Elizalde:1993ew}
\begin{eqnarray}
\begin{split}
&{ \beta  }_{ { \lambda  } }=\frac { 3{ \lambda  }^{ 2 } }{ { \left( 4\pi  \right)  }^{ 2 } } ,\quad{ \beta  }_{ { \xi  } }=\frac { { \lambda \left( \xi -1/6 \right)  } }{ { \left( 4\pi  \right)  }^{ 2 } } ,\quad { \beta  }_{ { { m }^{ 2 } } }=\frac { { \lambda { m }^{ 2 } } }{ { \left( 4\pi  \right)  }^{ 2 } } ,\quad \gamma_{\phi} =0 \\
&{ \beta  }_{ { { \rho  }_{ \Lambda  }  } }=\frac { { m }^{  4 } }{ { 2\left( 4\pi  \right)  }^{ 2 } } ,{ \quad \beta  }_{ { \kappa  } }=\frac { { { m }^{ 2 }\left( \xi -1/6 \right)  } }{ { \left( 4\pi  \right)  }^{ 2 } } , \\
&{ \beta  }_{ { { a }_{ 1 } } }=\frac { { { \left( \xi -1/6 \right)  }^{ 2 } } }{ { 2\left( 4\pi  \right)  }^{ 2 } } ,\quad { \beta  }_{ { { a }_{ 2 } } }=\frac { 1 }{ { 120\left( 4\pi  \right)  }^{ 2 } } , \\
&{ \beta  }_{ { { a }_{ 3 } } }=-\frac { 1 }{ { 360\left( 4\pi  \right)  }^{ 2 } } ,\quad { \beta  }_{ { { a }_{ 4 } } }=-\frac { { \xi -1/5 } }{ { 6\left( 4\pi  \right)  }^{ 2 } } ,\quad { \beta  }_{ { { a }_{ 5 } } }=-\frac { \lambda  }{ { 12\left( 4\pi  \right)  }^{ 2 } } .
\end{split}
\end{eqnarray}
where these one-loop $\beta$ functions corresponds to the logarithmic terms of Eq.~(\ref{eq:tutdfgdfsdg}).
Therefore, the vacuum energy of the effective potential 
does not run with any renormalization scale even in curved spacetime, and therefore, 
we need to clarify the physical interpretation of the RG running effects of the cosmological constant.

Now, let us reconsider the physical interpretation of the RG running effects of the cosmological constant
by considering the running gauge coupling of quantum electrodynamics (QED).
For simplicity, we consider the vacuum polarization diagrams in QED at the electron.
The one-loop contribution of the vacuum polarization diagrams ${ \Pi  }_{ 2 }^{\mu\nu}\left( { p }^{ 2 } \right)$ can be expressed as
\begin{equation}
{ \Pi  }_{ 2 }^{\mu\nu}\left( { p }^{ 2 } \right)={ e }^{ 2 }\int { \frac { { d }^{ 4 }k }{ { \left( 2\pi  \right)  }^{ 4 } }  } 
\frac { i }{ { \left( p-k \right)  }^{ 2 }-{ m }_{ e }^{ 2 } } \frac { i }{ { k }^{ 2 }-{ m }_{ e }^{ 2 } } {\rm Tr}
\left[ { \gamma  }^{ \mu  }\left( \Slash{k}-\Slash{p}+{ m }_{ e } \right) { \gamma  }^{ \nu  }\left( \Slash{k}+{ m }_{ e } \right)  \right],
\end{equation}
where $m_{e}$ is the electron mass and $p$ is the external momentum.
By using the dimensional regularization, we obtain the following expression
\begin{equation}
{ \Pi  }_{ 2 }^{\mu\nu}\left( { p }^{ 2 } \right)=i\frac { { e }^{ 2 } }{ 2{ \pi  }^{ 2 } } \left( { p }_{ \mu  }{ p }_{ \nu  }-{ p }^{ 2 }{ g }_{ \mu \nu  } \right) \left[ \frac { 1 }{ 6\epsilon   } -\frac { \gamma  }{ 6 } +\int _{ 0 }^{ 1 }{ dxx\left( 1-x \right) \ln { \frac {  4\pi { \mu  }^{ 2 }  }{ { m }_{ e }^{ 2 }+{ p }^{ 2 }x\left( 1-x \right)}}}  \right] ,
\end{equation}
where $\gamma$ is Euler's constant and $\epsilon $ is the renormalization parameter 
which are introduced by the dimensional regularization.
Here, we define ${ \Pi  }_{ 2 }\left( { p }^{ 2 } \right)$ as the following
\begin{equation}
{ \Pi  }_{ 2 }\left( { p }^{ 2 } \right)=\frac { 1}{ 2{ \pi  }^{ 2 } }\int _{ 0 }^{ 1 }{ dxx\left( 1-x \right) \ln { \frac {  4\pi { \mu  }^{ 2 }  }{ { m }_{ e }^{ 2 }+{ p }^{ 2 }x\left( 1-x \right)}}} .
\end{equation}
We can derive the effective Coulomb potential of the Fourier transform as follows
\begin{equation}
V_{\rm eff}\left( { p }^{ 2 } \right) ={ e }^{ 2 }\frac { 1-{ e }^{ 2 }{ \Pi  }_{ 2 }\left( { p }^{ 2 } \right) }{ { p }^{ 2 } } .
\end{equation}
If we define the gauge coupling $e$ at the low-energy scale $p_{0}$ as $V_{\rm eff}\left(p_{0}\right)=e^{2}/p^{2}_{0}$
and remove the divergence by using 
the counterterm, we can obtain the effective Coulomb potential as follows
\begin{equation}
V_{\rm eff}\left( { p }^{ 2 } \right) =\frac { { e }^{ 2 } }{ { p }^{ 2 } } \left\{ 1+\frac { { e }^{ 2 } }{ 2{ \pi  }^{ 2 } } \int _{ 0 }^{ 1 }{ dxx\left( 1-x \right) \ln { \left( \frac { { m }_{ e }^{2}+p^{ 2 }x\left( 1-x \right)  }{  { m }_{ e }^{2}+p_{0}^{ 2 }x\left( 1-x \right)  }  \right)  }  }  \right\} ,
\end{equation}
where we can safely remove the $\epsilon$ and $\gamma$.
In the high-energy limit ($p\gg m_{e}$), we can simply drop the electron mass-term $m_{e}$ and 
write the effective Coulomb potential as follows
\begin{equation}
V_{\rm eff }\left( { p }^{ 2 } \right) =\frac { { e }^{ 2 } }{ { p }^{ 2 } } \left( 1+\frac { { e }^{ 2 } }{ 12{ \pi  }^{ 2 } } 
\ln { \frac { p^{ 2 }  }{ p_{0}^{ 2 }  } }\right).
\end{equation}
Therefore, the RG running effects of the gauge coupling $e$ can be estimated as the following
\begin{eqnarray}
V_{\rm eff}\left( { p }^{ 2 } \right)=\frac { { e }^{ 2 } }{ { p }^{ 2 } } 
\left[ 1+\frac { { e }^{ 2 } }{ 12{ \pi  }^{ 2 } } \ln { \frac { p^{ 2 }  }{ p_{0}^{ 2 }  }}+
\left(\frac { { e }^{ 2 } }{ 12{ \pi  }^{ 2 } } \ln { \frac { p^{ 2 }  }{ p_{0}^{ 2 }  }}\right)^{2}+\cdots \right]
=\frac{1}{p^{2}}\left[ \frac{e^{2}}{1-\frac { { e }^{ 2 } }{ 12{ \pi  }^{ 2 } } \ln { \frac { p^{ 2 }  }{ p_{0}^{ 2 }  }}}\right]
\end{eqnarray}
The RG running coupling including the one-loop 1PI graphs is given by
\begin{equation}
e^{2}\left(p^{2}\right)= \frac{e^{2}}{1-\frac { { e }^{ 2 } }
{ 12{ \pi  }^{ 2 } } \ln { \frac { p^{ 2 }  }{ p_{0}^{ 2 }  }}}.
\end{equation}
On the other hand, the renormalization group equation (RGE) 
comes from the condition that the effective Coulomb potential is independent of the arbitrary scale $p_{0}$
\begin{equation}
\frac { d }{ d\ln { { p }_{ 0 }^{ 2 } }  }V_{\rm eff}\left( { p }^{ 2 } \right)  =0,
\end{equation}
where the RG running gauge coupling $e^{2}\left(p^{2}\right)$ offset the scale dependence of $p_{0}$ 
in the effective Coulomb potential $V_{\rm eff}\left( { p }^{ 2 } \right)$.
Therefore, the one-loop renormalization group equation is given by
\begin{equation}
\frac { d }{ d\ln { { p }_{ 0 }^{ 2 } }  } e\left( { p }^{ 2 } \right) =\frac { e^{3}\left( { p }^{ 2 } \right)  }{ 24{ \pi  }^{ 2 } } .
\end{equation}
By using the initial condition $e\left( { p }_{0}^{ 2 } \right)=e$, we can obtain the same result as the following
\begin{equation}
e^{2}\left(p^{2}\right)= \frac{e^{2}}{1-\frac { { e }^{ 2 } }
{ 12{ \pi  }^{ 2 } } \ln { \frac { p^{ 2 }  }{ p_{0}^{ 2 }  }}}.
\end{equation}
Thus, the renormalization group efficiently gives information about the high-energy scale~\cite{Mele:2006ji}.
The scale independence of the effective potential is theoretical requirement, but
the RG running of the coupling constant has definitely physical significance.

However, the physical meaning of the RG running is not simple in the case of the effective scalar potential. 
For simplicity, we consider the one-loop effective potential of massless scalar field theory as follows
\begin{eqnarray}
V_{\rm eff}\left( \phi  \right) =\frac{\lambda}{4}\phi^{4}
+\frac{9\lambda^{2}}{64\pi^{2}}\phi^{4}\left( \ln {\frac { 3\lambda \phi^{2} }{ { \mu  }^{ 2 } }} - \frac{3}{2} \right),
\end{eqnarray}
where we can simply absorb the factor of $-3/2$ into the definition of $\mu$
and drop this factor for simplicity. Because the renormalization scale $\mu$ is arbitrary parameter,
the self-coupling constant $\lambda$ and the scalar field $\phi$ must depend on $\mu$, but 
the effective scalar potential $V_{\rm eff}\left( \phi  \right)$ must be independent on the scale $\mu$.
The renormalization group equation (RGE) is derived from the scale independence of $V_{\rm eff}\left( \phi  \right)$
and given by  
\begin{equation}
\left( \mu \frac { \partial  }{ \partial \mu  } +{ \beta  }_{ { \lambda  } }\frac { \partial  }{ \partial { \lambda  }} -\gamma_{\phi} \phi \frac { \partial  }{ \partial \phi  }  \right)V_{\rm eff}\left(\phi \right) =0,
\end{equation}
where the $\beta$ function of ${ \lambda  }$ and the $\gamma$ function are defined by
\begin{equation}
{ \beta  }_{ \lambda  }\equiv  \frac { d\lambda  }{ d\ln{\mu}  }, \quad
\gamma_{\phi} \phi \equiv - \frac { d\phi  }{ d\ln{\mu}  } ,
\end{equation}
where self-coupling term $\lambda$ is defined at the scale $\mu$.
For convenience, we introduce the variable $t\equiv \ln { \left( \phi /\mu  \right)  } $ and obtain 
\begin{equation}
 \mu \frac { \partial  }{ \partial \mu  } =-\left(1+\gamma_{\phi} \right)\frac { \partial  }{ \partial t  }.
\end{equation}
Thus, the RGE is transformed as
\begin{equation}
\left( -\frac { \partial  }{ \partial t } +{ \beta }_{ { \lambda  } }'\frac { \partial  }{ \partial { \lambda  }} -\gamma'_{\phi}\phi \frac { \partial  }{ \partial \phi  }  \right)V_{\rm eff}\left(\phi \right) =0,
\end{equation}
where we define ${ \beta }_{ { \lambda  } }'\equiv \beta_{ \lambda  }/\left(1+\gamma_{\phi}\right)$ and 
$\gamma'_{\phi} \equiv \gamma_{\phi} /\left(1+\gamma_{\phi}\right)$ and we assume the simple form of 
the effective potential as
\begin{equation}
V_{\rm eff}\left( \phi  \right) =\frac{F\left(\lambda,t\right)}{4}\phi^{4}.
\end{equation}
Thus, the RGE of $F\left(\lambda,t\right)$ is written as
\begin{equation}
\left( -\frac { \partial  }{ \partial t } +{ \beta }_{ { \lambda  } }'\frac { \partial  }{ \partial { \lambda  }} -4\gamma'_{\phi} \right)
F\left(\lambda,t\right)=0.
\end{equation}
The solution of the RGE can be given as the following
\begin{equation}
F\left(\lambda,t\right)=\lambda\left(t\right) \exp
\left\{ -4\int _{ 0 }^{ t }{ ds\gamma'_{\phi}   \left[ \lambda \left( s \right)  \right]  }  \right\} ,
\end{equation}
where the running self-coupling constant $\lambda \left(t\right)$ is defined by
\begin{equation}
\frac{d\lambda\left(t\right)}{dt}={ \beta }_{ { \lambda  } }', \quad
\lambda\left(0\right)\equiv \lambda.
\end{equation}
In the one-loop order, we can take $\beta'_{ { \lambda  } }=\beta_{ { \lambda  } }$ and $\gamma'_{\phi}=\gamma_{\phi}$.
Furthermore, there is no wave function renormalization at one-loop order in the scalar field theory, and 
we can take the $\gamma$ function as $\gamma_{\phi}=0$.
Therefore, we obtain the RG improved effective potential as follows:
\begin{equation}
V_{\rm eff}\left(\phi\right)=\frac{1}{4}\lambda\left(t\right)\phi^{4},
\end{equation}
where we recall $t\equiv \ln { \left( \phi /\mu  \right)  } $ and
the running self-coupling $\lambda \left(t\right)$ is determined by 
the one-loop $\beta$ function is given as
\begin{equation}
\beta_{ { \lambda  } }\left(\lambda\right)=\frac{9\lambda^{2}}{16\pi^{2}}+\mathcal{O}\left(\lambda^{3}\right)+\cdots.
\end{equation}
By using the above result, we obtain the one-loop effective potential of massless scalar field theory as
\begin{eqnarray}
V_{\rm eff}\left(\phi\right)=\frac{\frac{1}{4}\lambda \phi^{4}}{1-\frac{9\lambda^{2}}{16\pi^{2}}\ln{\left(\frac{\phi^{2}}{\mu^{2}}\right)}}.
\end{eqnarray}
By expanding out the denominator, we can obtain the same form of the one-loop effective potential 
as the following
\begin{eqnarray}
V_{\rm eff}\left( \phi  \right) =\frac{\lambda}{4}\phi^{4}
+\frac{9\lambda^{2}}{64\pi^{2}}\phi^{4}\ln{\left(\frac{\phi^{2}}{\mu^{2}}\right)}+\cdots.
\end{eqnarray}
Here, we compare the effective scalar potential and the effective Coulomb potential as
\begin{eqnarray}
V_{\rm eff}\left( \phi  \right) =\frac{\lambda}{4}\phi^{4}
\left(1+\frac{9\lambda }{16\pi^{2}}\phi^{4}\ln{\frac{\phi^{2}}{\mu^{2}} }\right)  \Longleftrightarrow 
V_{\rm eff }\left( { p }^{ 2 } \right) =\frac { { e }^{ 2 } }{ { p }^{ 2 } } \left( 1+\frac { { e }^{ 2 } }{ 12{ \pi  }^{ 2 } } 
\ln { \frac { p^{ 2 }  }{ p_{0}^{ 2 }  } }\right) ,
\end{eqnarray}
where the self-coupling $\lambda$ is defined by the observation at the scale $\mu$ and 
the gauge coupling $e$ is defined by the observation at the momentum $p_{0}$.
The renormalization-scale invariance corresponds to the independence of
the initial and observational conditions against the formulation of the effective potential.
Next, we compare these in the context of the RGE as follows
\begin{eqnarray}
V_{\rm eff}\left( \phi  \right) =\frac{\lambda\left(\phi\right)}{4}\phi^{4} \Longleftrightarrow 
V_{\rm eff }\left( { p }^{ 2 } \right) =\frac { { e }^{ 2 }\left(p^{2}\right) }{ { p }^{ 2 } } ,
\end{eqnarray}
where the running self-coupling $\lambda\left(\phi\right)$ depend on the scalar field $\phi$ and 
the running gauge coupling ${ e }\left(p^{2}\right)$ depend on the momentum $p^{2}$. 
It is clear that the scalar field $\phi$ corresponds to the momentum $p^{2}$ on the effective Coulomb potential,
i.e the running coupling on the effective potential actually run by the scalar field $\phi$.

Finally, we consider the one-loop effective potential of the massive scalar field given by Eq.~(\ref{eq:tutusssdg}) in Minkowski spacetime.
If we use the running coupling of the RGE, we can obtain the RG improved effective potential as follows
\begin{eqnarray}
V_{\rm eff}\left( \phi  \right) =\frac{1}{2}m^{2}\left(t  \right)\phi^{2}+\frac{\lambda\left( t \right)}{4}\phi^{4}
+{ \rho  }_{ \Lambda  }\left( t  \right),
\end{eqnarray}
where we introduced the variable as $t=\ln \left({ { m }^{ 2 }+3\lambda \phi^{2} }/{ { \mu  }^{ 2 } }\right)$.
In the large field limit ($\phi\gg m$), we can assume $t\approx \ln \left({ \phi^{2} }/{ { \mu  }^{ 2 } }\right)$
and therefore, the running cosmological constant ${ \rho  }_{ \Lambda  }\left( t  \right)$ depend on the 
scalar field $\phi$. 
Therefore, the quantum vacuum energy density 
only change by the difference of the scalar field $\phi$ in Minkowski spacetime
(the case of the curved spacetime is discussed by Ref.\cite{Elizalde:1993ee,Elizalde:1993ew}).
The RG running effects can be described by the logarithmic terms, and therefore,
the running effects of the cosmological constant undoubtedly contribute the vacuum energy density.
However, the RG running of the cosmological constant has still some problems, e.g including decoupling effects~\cite{Gorbar:2002pw}
and does not give a direct proof of the dynamical vacuum energy ${ \rho  }_{ \rm vacuum } \simeq m^{2}H^{2}$ in literature.
However, as previously discussed, 
the vacuum energy density of ${ \rho  }_{ \rm vacuum } \simeq m^{2}H^{2}$ can appear 
as quantum particle production effects of the curved background rather than the RG running of the cosmological constant, and therefore, 
give a non-negligible impact on the observed Universe.

\section{Conclusion and Summary}
\label{sec:conclusion}
The current vacuum energy density observed as dark energy 
${ \rho }_{ \rm dark }\simeq 2.5\times10^{-47}\ {\rm GeV^{4}}$ is extremely small and 
raises serious problems for the theoretical physics.
In this work, we have reconsidered the RG running effects of the cosmological constant and
investigated the renormalized vacuum energy density in curved spacetime. 
The RG running effects of the cosmological constant undoubtedly contribute to the vacuum energy density,
but there are still some theoretical problems.
By adopting the method discussed by Bunch, Birrell and
Davies~\cite{bunch1978quantum,birrell1978application,Bunch:1980vc}, however, 
we have shown that the dynamical vacuum energy density described by ${ \rho }_{ \rm vacuum }\simeq m^{2}H^{2}$ appears as 
quantum particle production effects in the curved background rather than the RG running effects of cosmological constant.
This dynamical vacuum energy corresponds to the vacuum field fluctuation ${ \left< { \delta \phi }^{ 2 } \right> } $ and 
provides phenomenological contributions to the current Universe.
From the current cosmological
observations we have obtained the upper bound on the mass of the
background scalar field to be $m \lesssim M_{\rm Pl}$.

\acknowledgments 
We thank the referee for many thoughtful comments and suggestions.
This work is supported in part by MEXT KAKENHI Nos. JP15H05889,  JP16H00877, and JP17H01131 (K.K.), and JSPS KAKENHI No.~26247042 (K.K.).

\appendix
\section{Geometrical tensors in FLRW metric}
In the FLRW metric, 
the Ricci tensor and the Ricci scalar are given as follows~\cite{birrell1984quantum}
\begin{eqnarray}
\begin{split}
R_{00}&=\frac { 3 }{ 2 }D',\quad R_{11}=-\frac { 1 }{ 2 }\left(D'+D^{2}\right),
\quad R=\frac{3}{C}\left({ D' +\frac { 1 }{ 2 }  { D }^{ 2 } } \right),\\
G_{00}&=-\frac { 3 }{ 4 }D^{2},\quad G_{ii}=D'+\frac { 1}{ 4 }D^{2}, \\
H_{00}^{\left(1\right)}&=\frac { 9 }{ C }\left( \frac { 1 }{ 2 }D'^{2}-D''D +\frac{3}{8}D^{4} \right), \\
H_{ii}^{\left(1\right)}&=\frac { 3 }{ C }\left( 2D'''-D''D+\frac { 1 }{ 2 }D'^{2}-3D'D^{2} +\frac{3}{8}D^{4} \right).
\end{split}
\end{eqnarray}

\bibliographystyle{JHEP}
\bibliography{vacuum}

\providecommand{\href}[2]{#2}\begingroup\raggedright\begin{thebibliography}{100}

\bibitem{Knop:2003iy}
{\scshape Supernova Cosmology Project} collaboration, R.~A. Knop et~al.,
  \emph{{New constraints on Omega(M), Omega(lambda), and w from an independent
  set of eleven high-redshift supernovae observed with HST}},
  \href{http://dx.doi.org/10.1086/378560}{\emph{Astrophys. J.} {\bf 598} (2003)
  102}, [\href{http://arxiv.org/abs/astro-ph/0309368}{{\tt astro-ph/0309368}}].

\bibitem{Riess:2004nr}
{\scshape Supernova Search Team} collaboration, A.~G. Riess et~al., \emph{{Type
  Ia supernova discoveries at z > 1 from the Hubble Space Telescope: Evidence
  for past deceleration and constraints on dark energy evolution}},
  \href{http://dx.doi.org/10.1086/383612}{\emph{Astrophys. J.} {\bf 607} (2004)
  665--687}, [\href{http://arxiv.org/abs/astro-ph/0402512}{{\tt
  astro-ph/0402512}}].

\bibitem{Spergel:2006hy}
{\scshape WMAP} collaboration, D.~N. Spergel et~al., \emph{{Wilkinson Microwave
  Anisotropy Probe (WMAP) three year results: implications for cosmology}},
  \href{http://dx.doi.org/10.1086/513700}{\emph{Astrophys. J. Suppl.} {\bf 170}
  (2007) 377}, [\href{http://arxiv.org/abs/astro-ph/0603449}{{\tt
  astro-ph/0603449}}].

\bibitem{0067-0049-192-2-18}
E.~Komatsu, K.~M. Smith, J.~Dunkley, C.~L. Bennett, B.~Gold, G.~Hinshaw et~al.,
  \emph{Seven-year wilkinson microwave anisotropy probe (wmap) observations:
  Cosmological interpretation}, {\emph{The Astrophysical Journal Supplement
  Series} {\bf 192} (2011) 18}.

\bibitem{Ade:2013zuv}
{\scshape Planck} collaboration, P.~A.~R. Ade et~al., \emph{{Planck 2013
  results. XVI. Cosmological parameters}},
  \href{http://dx.doi.org/10.1051/0004-6361/201321591}{\emph{Astron.
  Astrophys.} {\bf 571} (2014) A16},
  [\href{http://arxiv.org/abs/1303.5076}{{\tt 1303.5076}}].

\bibitem{Weinberg:1988cp}
S.~Weinberg, \emph{{The Cosmological Constant Problem}},
  \href{http://dx.doi.org/10.1103/RevModPhys.61.1}{\emph{Rev. Mod. Phys.} {\bf
  61} (1989) 1--23}.

\bibitem{Sahni:1999gb}
V.~Sahni and A.~A. Starobinsky, \emph{{The Case for a positive cosmological
  Lambda term}}, \href{http://dx.doi.org/10.1142/S0218271800000542}{\emph{Int.
  J. Mod. Phys.} {\bf D9} (2000) 373--444},
  [\href{http://arxiv.org/abs/astro-ph/9904398}{{\tt astro-ph/9904398}}].

\bibitem{Carroll:2000fy}
S.~M. Carroll, \emph{{The Cosmological constant}},
  \href{http://dx.doi.org/10.12942/lrr-2001-1}{\emph{Living Rev. Rel.} {\bf 4}
  (2001) 1}, [\href{http://arxiv.org/abs/astro-ph/0004075}{{\tt
  astro-ph/0004075}}].

\bibitem{Padmanabhan:2002ji}
T.~Padmanabhan, \emph{{Cosmological constant: The Weight of the vacuum}},
  \href{http://dx.doi.org/10.1016/S0370-1573(03)00120-0}{\emph{Phys. Rept.}
  {\bf 380} (2003) 235--320}, [\href{http://arxiv.org/abs/hep-th/0212290}{{\tt
  hep-th/0212290}}].

\bibitem{Nojiri:2006ri}
S.~Nojiri and S.~D. Odintsov, \emph{{Introduction to modified gravity and
  gravitational alternative for dark energy}},
  \href{http://dx.doi.org/10.1142/S0219887807001928}{\emph{eConf} {\bf
  C0602061} (2006) 06}, [\href{http://arxiv.org/abs/hep-th/0601213}{{\tt
  hep-th/0601213}}].

\bibitem{Li:2011sd}
M.~Li, X.-D. Li, S.~Wang and Y.~Wang, \emph{{Dark Energy}},
  \href{http://dx.doi.org/10.1088/0253-6102/56/3/24}{\emph{Commun. Theor.
  Phys.} {\bf 56} (2011) 525--604}, [\href{http://arxiv.org/abs/1103.5870}{{\tt
  1103.5870}}].

\bibitem{Wang:2016och}
S.~Wang, Y.~Wang and M.~Li, \emph{{Holographic Dark Energy}},
  \href{http://arxiv.org/abs/1612.00345}{{\tt 1612.00345}}.

\bibitem{Martin:2012bt}
J.~Martin, \emph{{Everything You Always Wanted To Know About The Cosmological
  Constant Problem (But Were Afraid To Ask)}},
  \href{http://dx.doi.org/10.1016/j.crhy.2012.04.008}{\emph{Comptes Rendus
  Physique} {\bf 13} (2012) 566--665},
  [\href{http://arxiv.org/abs/1205.3365}{{\tt 1205.3365}}].

\bibitem{Peebles:2002gy}
P.~J.~E. Peebles and B.~Ratra, \emph{{The Cosmological constant and dark
  energy}}, \href{http://dx.doi.org/10.1103/RevModPhys.75.559}{\emph{Rev. Mod.
  Phys.} {\bf 75} (2003) 559--606},
  [\href{http://arxiv.org/abs/astro-ph/0207347}{{\tt astro-ph/0207347}}].

\bibitem{Copeland:2006wr}
E.~J. Copeland, M.~Sami and S.~Tsujikawa, \emph{{Dynamics of dark energy}},
  \href{http://dx.doi.org/10.1142/S021827180600942X}{\emph{Int. J. Mod. Phys.}
  {\bf D15} (2006) 1753--1936},
  [\href{http://arxiv.org/abs/hep-th/0603057}{{\tt hep-th/0603057}}].

\bibitem{Shapiro:2000dz}
I.~L. Shapiro and J.~Sola, \emph{{Scaling behavior of the cosmological
  constant: Interface between quantum field theory and cosmology}},
  \href{http://dx.doi.org/10.1088/1126-6708/2002/02/006}{\emph{JHEP} {\bf 02}
  (2002) 006}, [\href{http://arxiv.org/abs/hep-th/0012227}{{\tt
  hep-th/0012227}}].

\bibitem{Sola:2007sv}
J.~Sola, \emph{{Dark energy: A Quantum fossil from the inflationary
  Universe?}}, \href{http://dx.doi.org/10.1088/1751-8113/41/16/164066}{\emph{J.
  Phys.} {\bf A41} (2008) 164066}, [\href{http://arxiv.org/abs/0710.4151}{{\tt
  0710.4151}}].

\bibitem{Shapiro:2008yu}
I.~L. Shapiro and J.~Sola, \emph{{Can the cosmological 'constant' run? - It may
  run}},  \href{http://arxiv.org/abs/0808.0315}{{\tt 0808.0315}}.

\bibitem{Shapiro:2009dh}
I.~L. Shapiro and J.~Sola, \emph{{On the possible running of the cosmological
  'constant'}},
  \href{http://dx.doi.org/10.1016/j.physletb.2009.10.073}{\emph{Phys. Lett.}
  {\bf B682} (2009) 105--113}, [\href{http://arxiv.org/abs/0910.4925}{{\tt
  0910.4925}}].

\bibitem{Sola:2011qr}
J.~Sola, \emph{{Cosmologies with a time dependent vacuum}},
  \href{http://dx.doi.org/10.1088/1742-6596/283/1/012033}{\emph{J. Phys. Conf.
  Ser.} {\bf 283} (2011) 012033}, [\href{http://arxiv.org/abs/1102.1815}{{\tt
  1102.1815}}].

\bibitem{Sola:2013gha}
J.~Sola, \emph{{Cosmological constant and vacuum energy: old and new ideas}},
  \href{http://dx.doi.org/10.1088/1742-6596/453/1/012015}{\emph{J. Phys. Conf.
  Ser.} {\bf 453} (2013) 012015}, [\href{http://arxiv.org/abs/1306.1527}{{\tt
  1306.1527}}].

\bibitem{Shapiro:2003kv}
I.~L. Shapiro and J.~Sola, \emph{{Cosmological constant, renormalization group
  and Planck scale physics}},
  \href{http://dx.doi.org/10.1016/S0920-5632(03)02402-2}{\emph{Nucl. Phys.
  Proc. Suppl.} {\bf 127} (2004) 71--76},
  [\href{http://arxiv.org/abs/hep-ph/0305279}{{\tt hep-ph/0305279}}].

\bibitem{Shapiro:2003ui}
I.~L. Shapiro, J.~Sola, C.~Espana-Bonet and P.~Ruiz-Lapuente, \emph{{Variable
  cosmological constant as a Planck scale effect}},
  \href{http://dx.doi.org/10.1016/j.physletb.2003.09.016}{\emph{Phys. Lett.}
  {\bf B574} (2003) 149--155},
  [\href{http://arxiv.org/abs/astro-ph/0303306}{{\tt astro-ph/0303306}}].

\bibitem{EspanaBonet:2003vk}
C.~Espana-Bonet, P.~Ruiz-Lapuente, I.~L. Shapiro and J.~Sola, \emph{{Testing
  the running of the cosmological constant with type Ia supernovae at high z}},
  \href{http://dx.doi.org/10.1088/1475-7516/2004/02/006}{\emph{JCAP} {\bf 0402}
  (2004) 006}, [\href{http://arxiv.org/abs/hep-ph/0311171}{{\tt
  hep-ph/0311171}}].

\bibitem{Shapiro:2004is}
I.~L. Shapiro and J.~Sola, \emph{{A Friedmann-Lemaitre-Robertson-Walker
  cosmological model with running Lambda}},
  \href{http://arxiv.org/abs/astro-ph/0401015}{{\tt astro-ph/0401015}}.

\bibitem{Sola:2005et}
J.~Sola and H.~Stefancic, \emph{{Effective equation of state for dark energy:
  Mimicking quintessence and phantom energy through a variable lambda}},
  \href{http://dx.doi.org/10.1016/j.physletb.2005.08.051}{\emph{Phys. Lett.}
  {\bf B624} (2005) 147--157},
  [\href{http://arxiv.org/abs/astro-ph/0505133}{{\tt astro-ph/0505133}}].

\bibitem{Sola:2005nh}
J.~Sola and H.~Stefancic, \emph{{Dynamical dark energy or variable cosmological
  parameters?}}, \href{http://dx.doi.org/10.1142/S0217732306019554}{\emph{Mod.
  Phys. Lett.} {\bf A21} (2006) 479--494},
  [\href{http://arxiv.org/abs/astro-ph/0507110}{{\tt astro-ph/0507110}}].

\bibitem{Babic:2001vv}
A.~Babic, B.~Guberina, R.~Horvat and H.~Stefancic, \emph{{Renormalization group
  running of the cosmological constant and its implication for the Higgs boson
  mass in the standard model}},
  \href{http://dx.doi.org/10.1103/PhysRevD.65.085002}{\emph{Phys. Rev.} {\bf
  D65} (2002) 085002}, [\href{http://arxiv.org/abs/hep-ph/0111207}{{\tt
  hep-ph/0111207}}].

\bibitem{Babic:2004ev}
A.~Babic, B.~Guberina, R.~Horvat and H.~Stefancic, \emph{{Renormalization-group
  running cosmologies. A Scale-setting procedure}},
  \href{http://dx.doi.org/10.1103/PhysRevD.71.124041}{\emph{Phys. Rev.} {\bf
  D71} (2005) 124041}, [\href{http://arxiv.org/abs/astro-ph/0407572}{{\tt
  astro-ph/0407572}}].

\bibitem{Markkanen:2014poa}
T.~Markkanen, \emph{{Curvature induced running of the cosmological constant}},
  \href{http://dx.doi.org/10.1103/PhysRevD.91.124011}{\emph{Phys. Rev.} {\bf
  D91} (2015) 124011}, [\href{http://arxiv.org/abs/1412.3991}{{\tt
  1412.3991}}].

\bibitem{2012arXiv1204.1864C}
F.~E.~M. {Costa}, J.~A.~S. {Lima} and F.~A. {Oliveira}, \emph{{Decaying Vacuum
  Cosmology and its Scalar Field Description}}, {\emph{ArXiv e-prints} (Apr.,
  2012) }, [\href{http://arxiv.org/abs/1204.1864}{{\tt 1204.1864}}].

\bibitem{Sola:2014tta}
J.~Sola, \emph{{Vacuum energy and cosmological evolution}},
  \href{http://dx.doi.org/10.1063/1.4891113}{\emph{AIP Conf. Proc.} {\bf 1606}
  (2014) 19--37}, [\href{http://arxiv.org/abs/1402.7049}{{\tt 1402.7049}}].

\bibitem{Gomez-Valent:2014rxa}
A.~Gomez-Valent, J.~Sola and S.~Basilakos, \emph{{Dynamical vacuum energy in
  the expanding Universe confronted with observations: a dedicated study}},
  \href{http://dx.doi.org/10.1088/1475-7516/2015/01/004}{\emph{JCAP} {\bf 1501}
  (2015) 004}, [\href{http://arxiv.org/abs/1409.7048}{{\tt 1409.7048}}].

\bibitem{Sola:2015rra}
J.~Sola and A.~Gomez-Valent, \emph{{The $\bar{\Lambda}{\rm CDM}$ cosmology:
  From inflation to dark energy through running ?}},
  \href{http://dx.doi.org/10.1142/S0218271815410035}{\emph{Int. J. Mod. Phys.}
  {\bf D24} (2015) 1541003}, [\href{http://arxiv.org/abs/1501.03832}{{\tt
  1501.03832}}].

\bibitem{Sola:2015wwa}
J.~Sola, A.~Gomez-Valent and J.~de~Cruz~Perez, \emph{{Hints of dynamical vacuum
  energy in the expanding Universe}},
  \href{http://dx.doi.org/10.1088/2041-8205/811/1/L14}{\emph{Astrophys. J.}
  {\bf 811} (2015) L14}, [\href{http://arxiv.org/abs/1506.05793}{{\tt
  1506.05793}}].

\bibitem{Basilakos:2015vra}
S.~Basilakos and J.~Sola, \emph{{Growth index of matter perturbations in
  running vacuum models}},
  \href{http://dx.doi.org/10.1103/PhysRevD.92.123501}{\emph{Phys. Rev.} {\bf
  D92} (2015) 123501}, [\href{http://arxiv.org/abs/1509.06732}{{\tt
  1509.06732}}].

\bibitem{Sola:2016vis}
J.~Sola, \emph{{Running Vacuum in the Universe: current phenomenological
  status}},  \href{http://arxiv.org/abs/1601.01668}{{\tt 1601.01668}}.

\bibitem{Sola:2016jky}
J.~Sola, A.~Gomez-Valent and J.~d.~C. Perez, \emph{{First evidence of running
  cosmic vacuum: challenging the concordance model}},
  \href{http://arxiv.org/abs/1602.02103}{{\tt 1602.02103}}.

\bibitem{Fritzsch:2016ewd}
H.~Fritzsch, R.~C. Nunes and J.~Sola, \emph{{Running vacuum in the Universe and
  the time variation of the fundamental constants of Nature}},
  \href{http://arxiv.org/abs/1605.06104}{{\tt 1605.06104}}.

\bibitem{Sola:2016ecz}
J.~Sola, J.~d.~C. Perez, A.~Gomez-Valent and R.~C. Nunes, \emph{{Dynamical
  Vacuum against a rigid Cosmological Constant}},
  \href{http://arxiv.org/abs/1606.00450}{{\tt 1606.00450}}.

\bibitem{Sola:2016hnq}
J.~Sola, A.~Gomez-Valent and J.~d.~C. Perez, \emph{{Dynamical dark energy:
  scalar fields and running vacuum}},
  \href{http://arxiv.org/abs/1610.08965}{{\tt 1610.08965}}.

\bibitem{Sola:2016lle}
J.~Solà, \emph{{Cosmological constant vis-à-vis dynamical vacuum: Bold
  challenging the $\Lambda$CDM}},
  \href{http://dx.doi.org/10.1142/S0217751X16300350}{\emph{Int. J. Mod. Phys.}
  {\bf A31} (2016) 1630035}, [\href{http://arxiv.org/abs/1612.02449}{{\tt
  1612.02449}}].

\bibitem{Foot:2007wn}
R.~Foot, A.~Kobakhidze, K.~L. McDonald and R.~R. Volkas,
  \emph{{Renormalization-scale independence of the physical cosmological
  constant}},
  \href{http://dx.doi.org/10.1016/j.physletb.2008.05.029}{\emph{Phys. Lett.}
  {\bf B664} (2008) 199--200}, [\href{http://arxiv.org/abs/0712.3040}{{\tt
  0712.3040}}].

\bibitem{Ward:2009wq}
B.~F.~L. Ward, \emph{{On the Running of the Cosmological Constant in Quantum
  General Relativity}},
  \href{http://dx.doi.org/10.1142/S021773231003269X}{\emph{Mod. Phys. Lett.}
  {\bf A25} (2010) 607--610}, [\href{http://arxiv.org/abs/0908.1764}{{\tt
  0908.1764}}].

\bibitem{Ward:2014sla}
B.~F.~L. Ward, \emph{{Running of the Cosmological Constant and Estimate of its
  Value in Quantum General Relativity}},
  \href{http://dx.doi.org/10.1142/S0217732315400301}{\emph{Mod. Phys. Lett.}
  {\bf A30} (2015) 1540030}, [\href{http://arxiv.org/abs/1412.7417}{{\tt
  1412.7417}}].

\bibitem{Hamber:2013rb}
H.~W. Hamber and R.~Toriumi, \emph{{Inconsistencies from a Running Cosmological
  Constant}}, \href{http://dx.doi.org/10.1142/S0218271813300231}{\emph{Int. J.
  Mod. Phys.} {\bf D22} (2013) 1330023},
  [\href{http://arxiv.org/abs/1301.6259}{{\tt 1301.6259}}].

\bibitem{Mele:2006ji}
S.~Mele, \emph{{Measurements of the running of the electromagnetic coupling at
  LEP}},  in \emph{{Proceedings, 26th International Symposium on Physics in
  Collision (PIC 2006): Buzios, Brazil, July 6-9, 2006}}, 2006.
\newblock \href{http://arxiv.org/abs/hep-ex/0610037}{{\tt hep-ex/0610037}}.

\bibitem{bunch1978quantum}
T.~S. Bunch and P.~C. Davies, \emph{Quantum field theory in de sitter space:
  renormalization by point-splitting},  in \emph{Proceedings of the Royal
  Society of London A: Mathematical, Physical and Engineering Sciences},
  vol.~360, pp.~117--134, The Royal Society, 1978.

\bibitem{birrell1978application}
N.~Birrell, \emph{The application of adiabatic regularization to calculations
  of cosmological interest},  in \emph{Proceedings of the Royal Society of
  London A: Mathematical, Physical and Engineering Sciences}, vol.~361,
  pp.~513--526, The Royal Society, 1978.

\bibitem{Bunch:1980vc}
T.~S. Bunch, \emph{{ADIABATIC REGULARIZATION FOR SCALAR FIELDS WITH ARBITRARY
  COUPLING TO THE SCALAR CURVATURE}},
  \href{http://dx.doi.org/10.1088/0305-4470/13/4/022}{\emph{J. Phys.} {\bf A13}
  (1980) 1297--1310}.

\bibitem{Asorey:2012xq}
M.~Asorey, P.~M. Lavrov, B.~J. Ribeiro and I.~L. Shapiro, \emph{{Vacuum
  stress-tensor in SSB theories}},
  \href{http://dx.doi.org/10.1103/PhysRevD.85.104001}{\emph{Phys. Rev.} {\bf
  D85} (2012) 104001}, [\href{http://arxiv.org/abs/1202.4235}{{\tt
  1202.4235}}].

\bibitem{Maggiore:2010wr}
M.~Maggiore, \emph{{Zero-point quantum fluctuations and dark energy}},
  \href{http://dx.doi.org/10.1103/PhysRevD.83.063514}{\emph{Phys. Rev.} {\bf
  D83} (2011) 063514}, [\href{http://arxiv.org/abs/1004.1782}{{\tt
  1004.1782}}].

\bibitem{Bilic:2011zm}
N.~Bilic, \emph{{Vacuum fluctuations in a supersymmetric model in FRW
  spacetime}}, \href{http://dx.doi.org/10.1103/PhysRevD.83.105003}{\emph{Phys.
  Rev.} {\bf D83} (2011) 105003}, [\href{http://arxiv.org/abs/1104.1349}{{\tt
  1104.1349}}].

\bibitem{Bilic:2011rj}
N.~Bilic, S.~Domazet and B.~Guberina, \emph{{Vacuum fluctuations of the
  supersymmetric field in curved background}},
  \href{http://dx.doi.org/10.1016/j.physletb.2011.12.025}{\emph{Phys. Lett.}
  {\bf B707} (2012) 221--227}, [\href{http://arxiv.org/abs/1110.2393}{{\tt
  1110.2393}}].

\bibitem{Hollenstein:2011cz}
L.~Hollenstein, M.~Jaccard, M.~Maggiore and E.~Mitsou, \emph{{Zero-point
  quantum fluctuations in cosmology}},
  \href{http://dx.doi.org/10.1103/PhysRevD.85.124031}{\emph{Phys. Rev.} {\bf
  D85} (2012) 124031}, [\href{http://arxiv.org/abs/1111.5575}{{\tt
  1111.5575}}].

\bibitem{Hack:2013uyu}
T.-P. Hack, \emph{{The Lambda CDM-model in quantum field theory on curved
  spacetime and Dark Radiation}},  \href{http://arxiv.org/abs/1306.3074}{{\tt
  1306.3074}}.

\bibitem{Fredenhagen:2013vxa}
K.~Fredenhagen and T.-P. Hack, \emph{{Quantum field theory on curved spacetime
  and the standard cosmological model}},
  \href{http://dx.doi.org/10.1007/978-3-662-46422-9_6}{\emph{Lect. Notes Phys.}
  {\bf 899} (2015) 113--129}, [\href{http://arxiv.org/abs/1308.6773}{{\tt
  1308.6773}}].

\bibitem{Padmanabhan:2004qc}
T.~Padmanabhan, \emph{{Vacuum fluctuations of energy density can lead to the
  observed cosmological constant}},
  \href{http://dx.doi.org/10.1088/0264-9381/22/17/L01}{\emph{Class. Quant.
  Grav.} {\bf 22} (2005) L107--L110},
  [\href{http://arxiv.org/abs/hep-th/0406060}{{\tt hep-th/0406060}}].

\bibitem{Ford:1997hb}
L.~H. Ford, \emph{{Quantum field theory in curved space-time}},  in
  \emph{{Particles and fields. Proceedings, 9th Jorge Andre Swieca Summer
  School, Campos do Jordao, Brazil, February 16-28, 1997}}, pp.~345--388, 1997.
\newblock \href{http://arxiv.org/abs/gr-qc/9707062}{{\tt gr-qc/9707062}}.

\bibitem{Giudice:2013nak}
G.~F. Giudice, \emph{{Naturalness after LHC8}}, {\emph{PoS} {\bf EPS-HEP2013}
  (2013) 163}, [\href{http://arxiv.org/abs/1307.7879}{{\tt 1307.7879}}].

\bibitem{Farina:2013mla}
M.~Farina, D.~Pappadopulo and A.~Strumia, \emph{{A modified naturalness
  principle and its experimental tests}},
  \href{http://dx.doi.org/10.1007/JHEP08(2013)022}{\emph{JHEP} {\bf 08} (2013)
  022}, [\href{http://arxiv.org/abs/1303.7244}{{\tt 1303.7244}}].

\bibitem{Dine:2015xga}
M.~Dine, \emph{{Naturalness Under Stress}},
  \href{http://dx.doi.org/10.1146/annurev-nucl-102014-022053}{\emph{Ann. Rev.
  Nucl. Part. Sci.} {\bf 65} (2015) 43--62},
  [\href{http://arxiv.org/abs/1501.01035}{{\tt 1501.01035}}].

\bibitem{Matsui:2016cls}
H.~Matsui and Y.~Matsumoto, \emph{{Gravitational relaxation of electroweak
  hierarchy problem}},  \href{http://arxiv.org/abs/1608.08838}{{\tt
  1608.08838}}.

\bibitem{Ringwald:1987ui}
A.~Ringwald, \emph{{Evolution Equation for the Expectation Value of a Scalar
  Field in Spatially Flat Rw Universes}},
  \href{http://dx.doi.org/10.1016/S0003-4916(87)80027-1}{\emph{Annals Phys.}
  {\bf 177} (1987) 129}.

\bibitem{Fulling:1974pu}
S.~A. Fulling, L.~Parker and B.~L. Hu, \emph{{Conformal energy-momentum tensor
  in curved spacetime: Adiabatic regularization and renormalization}},
  \href{http://dx.doi.org/10.1103/PhysRevD.10.3905}{\emph{Phys. Rev.} {\bf D10}
  (1974) 3905--3924}.

\bibitem{Fulling:1974zr}
S.~A. Fulling and L.~Parker, \emph{{Renormalization in the theory of a
  quantized scalar field interacting with a robertson-walker spacetime}},
  \href{http://dx.doi.org/10.1016/0003-4916(74)90451-5}{\emph{Annals Phys.}
  {\bf 87} (1974) 176--204}.

\bibitem{Parker:1974qw}
L.~Parker and S.~A. Fulling, \emph{{Adiabatic regularization of the energy
  momentum tensor of a quantized field in homogeneous spaces}},
  \href{http://dx.doi.org/10.1103/PhysRevD.9.341}{\emph{Phys. Rev.} {\bf D9}
  (1974) 341--354}.

\bibitem{Anderson:1987yt}
P.~R. Anderson and L.~Parker, \emph{{Adiabatic Regularization in Closed
  Robertson-walker Universes}},
  \href{http://dx.doi.org/10.1103/PhysRevD.36.2963}{\emph{Phys. Rev.} {\bf D36}
  (1987) 2963}.

\bibitem{Paz:1988mt}
J.~P. Paz and F.~D. Mazzitelli, \emph{{Renormalized Evolution Equations for the
  Back Reaction Problem With a Selfinteracting Scalar Field}},
  \href{http://dx.doi.org/10.1103/PhysRevD.37.2170}{\emph{Phys. Rev.} {\bf D37}
  (1988) 2170--2181}.

\bibitem{Haro:2010zz}
J.~Haro, \emph{{Calculation of the renormalized two-point function by adiabatic
  regularization}},
  \href{http://dx.doi.org/10.1007/s11232-010-0123-2}{\emph{Theor. Math. Phys.}
  {\bf 165} (2010) 1490--1499}.

\bibitem{Haro:2010mx}
J.~Haro, \emph{{Topics in Quantum Field Theory in Curved Space}},
  \href{http://arxiv.org/abs/1011.4772}{{\tt 1011.4772}}.

\bibitem{Kohri:2016qqv}
K.~Kohri and H.~Matsui, \emph{{Electroweak Vacuum Instability and Renormalized
  Higgs Field Vacuum Fluctuations in the Inflationary Universe}},
  \href{http://arxiv.org/abs/1607.08133}{{\tt 1607.08133}}.

\bibitem{birrell1984quantum}
N.~D. Birrell and P.~C.~W. Davies, \emph{Quantum fields in curved space}.
\newblock No.~7. Cambridge university press, 1984.

\bibitem{Vilenkin:1982wt}
A.~Vilenkin and L.~H. Ford, \emph{{Gravitational Effects upon Cosmological
  Phase Transitions}},
  \href{http://dx.doi.org/10.1103/PhysRevD.26.1231}{\emph{Phys. Rev.} {\bf D26}
  (1982) 1231}.

\bibitem{Elias:2015yta}
M.~Elias and F.~D. Mazzitelli, \emph{{Ultraviolet cutoffs for quantum fields in
  cosmological spacetimes}},
  \href{http://dx.doi.org/10.1103/PhysRevD.91.124051}{\emph{Phys. Rev.} {\bf
  D91} (2015) 124051}, [\href{http://arxiv.org/abs/1504.02993}{{\tt
  1504.02993}}].

\bibitem{Habib:1999cs}
S.~Habib, C.~Molina-Paris and E.~Mottola, \emph{{Energy momentum tensor of
  particles created in an expanding universe}},
  \href{http://dx.doi.org/10.1103/PhysRevD.61.024010}{\emph{Phys. Rev.} {\bf
  D61} (2000) 024010}, [\href{http://arxiv.org/abs/gr-qc/9906120}{{\tt
  gr-qc/9906120}}].

\bibitem{Anderson:2000wx}
P.~R. Anderson, W.~Eaker, S.~Habib, C.~Molina-Paris and E.~Mottola,
  \emph{{Attractor states and infrared scaling in de Sitter space}},
  \href{http://dx.doi.org/10.1103/PhysRevD.62.124019}{\emph{Phys. Rev.} {\bf
  D62} (2000) 124019}, [\href{http://arxiv.org/abs/gr-qc/0005102}{{\tt
  gr-qc/0005102}}].

\bibitem{Anderson:2013ila}
P.~R. Anderson and E.~Mottola, \emph{{Instability of global de Sitter space to
  particle creation}},
  \href{http://dx.doi.org/10.1103/PhysRevD.89.104038}{\emph{Phys. Rev.} {\bf
  D89} (2014) 104038}, [\href{http://arxiv.org/abs/1310.0030}{{\tt
  1310.0030}}].

\bibitem{Anderson:2013zia}
P.~R. Anderson and E.~Mottola, \emph{{Quantum vacuum instability of ?eternal?
  de Sitter space}},
  \href{http://dx.doi.org/10.1103/PhysRevD.89.104039}{\emph{Phys. Rev.} {\bf
  D89} (2014) 104039}, [\href{http://arxiv.org/abs/1310.1963}{{\tt
  1310.1963}}].

\bibitem{Markkanen:2016aes}
T.~Markkanen and A.~Rajantie, \emph{{Massive scalar field evolution in de
  Sitter}},  \href{http://arxiv.org/abs/1607.00334}{{\tt 1607.00334}}.

\bibitem{Markkanen:2016jhg}
T.~Markkanen, \emph{{Decoherence Can Relax Cosmic Acceleration}},
  \href{http://dx.doi.org/10.1088/1475-7516/2016/11/026}{\emph{JCAP} {\bf 1611}
  (2016) 026}, [\href{http://arxiv.org/abs/1609.01738}{{\tt 1609.01738}}].

\bibitem{Christensen:1977jc}
S.~M. Christensen and S.~A. Fulling, \emph{{Trace Anomalies and the Hawking
  Effect}}, \href{http://dx.doi.org/10.1103/PhysRevD.15.2088}{\emph{Phys. Rev.}
  {\bf D15} (1977) 2088--2104}.

\bibitem{Candelas:1980zt}
P.~Candelas, \emph{{Vacuum Polarization in Schwarzschild Space-Time}},
  \href{http://dx.doi.org/10.1103/PhysRevD.21.2185}{\emph{Phys. Rev.} {\bf D21}
  (1980) 2185--2202}.

\bibitem{Christensen:1976vb}
S.~M. Christensen, \emph{{Vacuum Expectation Value of the Stress Tensor in an
  Arbitrary Curved Background: The Covariant Point Separation Method}},
  \href{http://dx.doi.org/10.1103/PhysRevD.14.2490}{\emph{Phys. Rev.} {\bf D14}
  (1976) 2490--2501}.

\bibitem{Christensen:1978yd}
S.~M. Christensen, \emph{{Regularization, Renormalization, and Covariant
  Geodesic Point Separation}},
  \href{http://dx.doi.org/10.1103/PhysRevD.17.946}{\emph{Phys. Rev.} {\bf D17}
  (1978) 946--963}.

\bibitem{Dvali:2010bf}
G.~Dvali and C.~Gomez, \emph{{Self-Completeness of Einstein Gravity}},
  \href{http://arxiv.org/abs/1005.3497}{{\tt 1005.3497}}.

\bibitem{Dvali:2010ue}
G.~Dvali, S.~Folkerts and C.~Germani, \emph{{Physics of Trans-Planckian
  Gravity}}, \href{http://dx.doi.org/10.1103/PhysRevD.84.024039}{\emph{Phys.
  Rev.} {\bf D84} (2011) 024039}, [\href{http://arxiv.org/abs/1006.0984}{{\tt
  1006.0984}}].

\bibitem{Elizalde:1993ee}
E.~Elizalde and S.~D. Odintsov, \emph{{Renormalization group improved effective
  potential for gauge theories in curved space-time}},
  \href{http://dx.doi.org/10.1016/0370-2693(93)91427-O}{\emph{Phys. Lett.} {\bf
  B303} (1993) 240--248}, [\href{http://arxiv.org/abs/hep-th/9302074}{{\tt
  hep-th/9302074}}].

\bibitem{Elizalde:1993ew}
E.~Elizalde and S.~D. Odintsov, \emph{{Renormalization group improved effective
  Lagrangian for interacting theories in curved space-time}},
  \href{http://dx.doi.org/10.1016/0370-2693(94)90464-2}{\emph{Phys. Lett.} {\bf
  B321} (1994) 199--204}, [\href{http://arxiv.org/abs/hep-th/9311087}{{\tt
  hep-th/9311087}}].

\bibitem{Elizalde:1993qh}
E.~Elizalde and S.~D. Odintsov, \emph{{Renormalization group improved effective
  potential for interacting theories with several mass scales in curved
  space-time}}, \href{http://dx.doi.org/10.1007/BF01957780}{\emph{Z. Phys.}
  {\bf C64} (1994) 699--708}, [\href{http://arxiv.org/abs/hep-th/9401057}{{\tt
  hep-th/9401057}}].

\bibitem{Elizalde:1994gv}
E.~Elizalde, S.~D. Odintsov and A.~Romeo, \emph{{Improved effective potential
  in curved space-time and quantum matter, higher derivative gravity theory}},
  \href{http://dx.doi.org/10.1103/PhysRevD.51.1680}{\emph{Phys. Rev.} {\bf D51}
  (1995) 1680--1691}, [\href{http://arxiv.org/abs/hep-th/9410113}{{\tt
  hep-th/9410113}}].

\bibitem{Elizalde:1995at}
E.~Elizalde, C.~O. Lousto, S.~D. Odintsov and A.~Romeo, \emph{{GUTs in curved
  space-time: Running gravitational constants, Newtonian potential and the
  quantum corrected gravitational equations}},
  \href{http://dx.doi.org/10.1103/PhysRevD.52.2202}{\emph{Phys. Rev.} {\bf D52}
  (1995) 2202--2213}, [\href{http://arxiv.org/abs/hep-th/9504014}{{\tt
  hep-th/9504014}}].

\bibitem{parker2009quantum}
L.~Parker and D.~Toms, \emph{Quantum field theory in curved spacetime:
  quantized fields and gravity}.
\newblock Cambridge University Press, 2009.

\bibitem{birrell1982p}
N.~Birrell, \emph{P. Davies Quantum Fields in Curved Space}.
\newblock Cambridge University Press, 1982.

\bibitem{fulling1989aspects}
S.~A. Fulling, \emph{Aspects of quantum field theory in curved spacetime},
  vol.~17.
\newblock Cambridge University Press, 1989.

\bibitem{DeWitt:1975ys}
B.~S. DeWitt, \emph{{Quantum Field Theory in Curved Space-Time}},
  \href{http://dx.doi.org/10.1016/0370-1573(75)90051-4}{\emph{Phys. Rept.} {\bf
  19} (1975) 295--357}.

\bibitem{Buchbinder:1992rb}
I.~L. Buchbinder, S.~D. Odintsov and I.~L. Shapiro, \emph{{Effective action in
  quantum gravity}}.
\newblock 1992.

\bibitem{Bunch:1979uk}
T.~S. Bunch and L.~Parker, \emph{{Feynman Propagator in Curved Space-Time: A
  Momentum Space Representation}},
  \href{http://dx.doi.org/10.1103/PhysRevD.20.2499}{\emph{Phys. Rev.} {\bf D20}
  (1979) 2499--2510}.

\bibitem{Shapiro:2008sf}
I.~L. Shapiro, \emph{{Effective Action of Vacuum: Semiclassical Approach}},
  \href{http://dx.doi.org/10.1088/0264-9381/25/10/103001}{\emph{Class. Quant.
  Grav.} {\bf 25} (2008) 103001}, [\href{http://arxiv.org/abs/0801.0216}{{\tt
  0801.0216}}].

\bibitem{Gorbar:2002pw}
E.~V. Gorbar and I.~L. Shapiro, \emph{{Renormalization group and decoupling in
  curved space}},
  \href{http://dx.doi.org/10.1088/1126-6708/2003/02/021}{\emph{JHEP} {\bf 02}
  (2003) 021}, [\href{http://arxiv.org/abs/hep-ph/0210388}{{\tt
  hep-ph/0210388}}].

\end{thebibliography}\endgroup
\end{document}